\documentclass[12pt,preprint]{aastex}

\shorttitle{DAO Liquid Crystal Spectropolarimeter}
\shortauthors{Monin et al.}

\begin{document}

\title{An Inexpensive  Liquid Crystal Spectropolarimeter for the Dominion Astrophysical Observatory Plaskett Telescope }

\author{D. Monin,
       D. Bohlender,
       T. Hardy,
       L. Saddlemyer,
       and M. Fletcher
       }
\affil{Herzberg Institute of Astrophysics,\\
        National Research Council of Canada,\\
        5071 West Saanich Road, Victoria BC  V9E 2E7\\
       }

\begin{abstract}
A new, inexpensive polarimetric unit has been constructed for the Dominion Astrophysical Observatory (DAO) 1.8-m Plaskett
telescope. It is implemented as a plug-in module for the telescope's existing Cassegrain
spectrograph, and enables medium resolution ($R\approx10,000$) circular
spectropolarimetry of point sources.
A dual-beam design together with fast switching of the wave plate at rates up to
100\,Hz, and synchronized with charge shuffling on the CCD, is used to
significantly reduce instrumental effects and achieve high-precision
spectropolarimetric measurements for a very low cost.
The instrument is optimized to work in the wavelength range
$4700 - 5300$\,\AA\, to simultaneously detect polarization signals in the H$\beta$ line
as well as nearby metallic lines.
In this paper we describe the technical details of the instrument, our observing strategy and data reduction
techniques, and present tests of its scientific performance.
\end{abstract}

\keywords{Stars -- Astronomical Instrumentation -- Data Analysis and Techniques}

\section{Introduction}

Spectropolarimetry has been used for decades to detect 
magnetic fields in stars throughout the HR diagram \citep{donati09}.  
Our particular interest in this paper, however, is in instrumentation that enables the measurement of strong, large-scale magnetic fields in the chemically peculiar stars (Ap/Bp) of the upper main sequence.
Some existing polarimeters permit
high-resolution (defined here as $R > 40,000$), high-precision spectropolarimetry in metallic lines of such stars
(for example, MuSiCoS and NARVAL on the 2\,m Bernard Lyot Telescope
at Pic du Midi, and ESPaDOnS on the 3.6\,m Canada-France-Hawaii Telescope)
but hydrogen lines, due to their large width, provide much lower accuracy.
Low-resolution ($R \approx 2,000$) instruments such as
FORS1 on the 8\,m Antu telescope of the VLT \citep{bagnulo02}
enable more precise spectropolarimetric measurements of hydrogen lines
but at a cost of providing such low contrast for the metallic lines that they can not be fully utilized.
Because of this,  high-resolution spectropolarimeters are generally used to measure magnetic fields in Ap and Bp stars using only
metallic (and perhaps helium) lines, while only hydrogen lines are measured with
low-resolution instruments.

However, for many years it has been known that a number of stars show quite significant discrepancies between
magnetic field strengths derived from hydrogen lines versus values found using metallic lines, but with different instruments
\citep[for example,][]{borra80,bychkov05}.
The differences can be from
a few hundred G to a few kG, or more than 10\% of the field strength since longitudinal magnetic fields in these stars can range from several hundred G to several kG.
As an example, in 52\,Her the average longitudinal field derived from metallic lines
is 1500\,G while the average field derived from hydrogen lines is 500\,G
\citep{bychkov05}!
Unfortunately, it is unclear what part of the observed differences might be due to details of the
instrumentation and data analysis techniques and what part may be produced by the very common and often very pronounced non-uniform horizontal and vertical abundance distributions of metals in the photospheres of the Ap and Bp stars.

In an attempt to resolve this long-standing question
we have investigated the possibility of constructing a low-cost, medium-resolution spectropolarimeter that provides circular polarization measurements of hydrogen
and metallic lines of Ap and Bp stars and hence enables the measurement of longitudinal magnetic fields with comparable precision using both diagnostics simultaneously.
The low cost was dictated by the limited operations and development budget for the DAO telescopes.
Despite this constraint, we set out to build an instrument that would let us measure stellar magnetic fields in relatively bright stars
with typical accuracies of about 100\,G and better than 50\,G for the brightest targets.
We will demonstrate later that such precisions are sufficient
to routinely detect discrepancies between hydrogen and metallic line longitudinal
field measurements and also show for the first time that for at least one object
these discrepancies do {\em not} have an instrumental origin.

A second science goal of the new  polarimeter was to provide us with the ability to carry out (in a reasonable length of time) a large,
sensitive search for globally-ordered magnetic fields in relatively faint
and possibly rapidly rotating upper main sequence peculiar stars observable
from the DAO.  For hundreds of such poorly studied 6th and 7th\,mag Ap and Bp stars we therefore needed to obtain field measurements with an accuracy on the order of 200\,G from observations acquired in two hours or less.
Such measurements, obtained many times over the rotation period of the star, are sufficient to confidently detect and characterize the large-scale magnetic fields observed in many upper main sequence peculiar magnetic stars.

In the remainder of this paper we describe the optical and mechanical design
of the spectropolarimeter {\em `dimaPol'}
that we have designed and constructed for use on the DAO's 1.8-m Plaskett telescope 
for a very low cost of only several thousand dollars
(including spare components and parts used only for testing).
The polarimeter module can be switched at a rate of up to 100\,Hz and is implemented as a plug-in module for the existing Cassegrain spectrograph on the DAO 1.8-m telescope.
We describe our data reduction technique that not only enabled the use of low cost, off-the-shelf components in the polarimeter but also provides near real-time conversion of the spectropolarimetric data to a magnetic field measurement.
A summary of tests we have carried out to investigate the optimal modulation rate of the polarimeter and the best observing strategy is provided, and we also present results of extensive new magnetic field measurements of well-known magnetic stars as well as non-magnetic standards and demonstrate how these new data compare very favourably to observations made with other generally much more costly spectropolarimeters.

\section{The DAO Liquid Crystal Spectropolarimeter}

The level of polarization observed in starlight is usually very weak. Numerous
instrumental effects can make these weak signals 
undetectable or lead to spurious detections.
Most astronomical spectropolarimeters  currently in operation use a dual-beam design
to deal with this issue.
The polarization of the input beam is transformed into
an intensity difference between the two output beams.
Since both spectra are taken under exactly the same conditions,
the polarization signal is independent of sky transparency, seeing, or slit
losses but not the instrumental systematics.
A polarization modulator is therefore used in many spectropolarimeters
to fight these possible instrumental effects by interchanging the two output beams so 
that they travel  exactly the same optical path through the instrument.
Usually the modulation is achieved by means of the mechanical rotation of an optical element
but this can be a relatively slow process.
Since at least two observations are now needed for each distinct configuration of the device and these are separated in time,
guiding errors can now come into play.
These can cause the intensity distribution of the input light beam
across the entrance aperture of the spectrograph to change from one observation
to the next which in turn can lead to a wavelength shift on the detector and
hence may result in a spurious polarization signal.
A change in brightness due to seeing and transparency variations
and different gains in the two instrument channels can also generate
a spurious signal \citep[see the discussion of this effect in][]{povel01}.
This can be a limiting factor for the sensitivity of a spectropolarimeter unless the instrument is fed
by an optical fiber (as in ESPaDOnS, for example) or some form of fast modulation is used.
The spectrograph currently permanently installed on the DAO Plaskett 1.8-m telescope is  
fed directly from the telescope's Cassegrain focus.
Because of this fact, as well as budget constraints, we confined our design considerations for a new DAO spectropolarimeter
to those that made use of a fast modulation option.

\begin{figure}[t!]
\includegraphics[angle=0,scale=1]{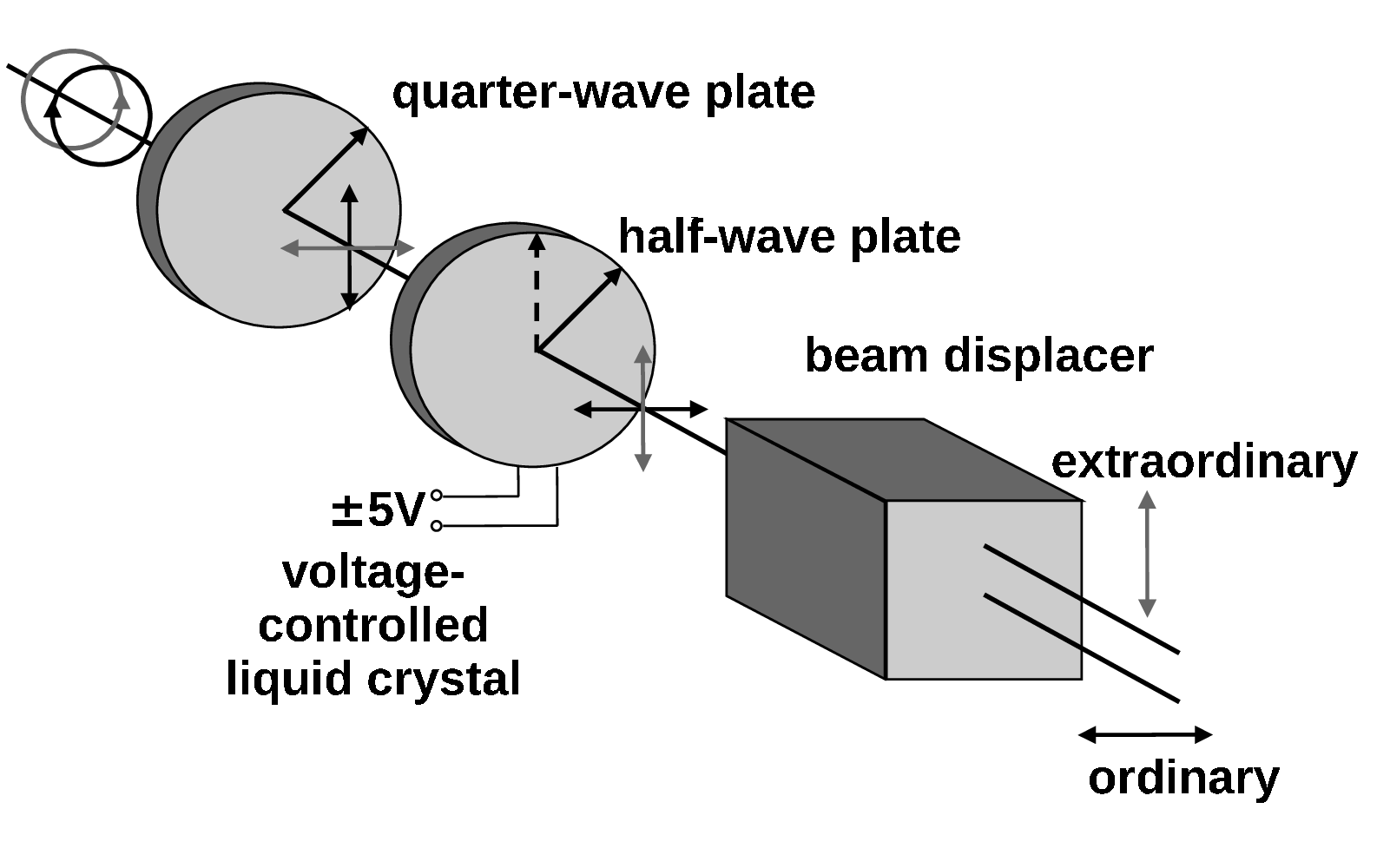}
\caption{A sketch of the optical layout of the polarimetric module.
\label{module_sketch}}
\end{figure}

{\em dimaPol} is installed in the DAO 1.8-m telescope's $f/18$ diverging beam immediately behind one of three
possible circular entrance apertures to the Cassegrain spectrograph.
The module consists of an achromatic quarter-wave plate,
a ferro-electric liquid crystal (FLC) half-wave plate,
a calcite beam displacer, and a mechanical shutter
(Figure~\ref{module_sketch}).
We  discuss the individual optical components in detail below.
The single-order grating spectrograph has 
an off-axis hyperbolic collimator mirror, a plane diffraction
grating, and a spherical camera mirror \citep{richardson68}.
An 1800\,lines\,mm$^{-1}$ grating provides a spectral resolution
of $R = 8,000$ to $15,000$ depending on the choice of entrance aperture.
The 1752$\times$532 SITe CCD with $15\,\mu$m pixels captures an approximately 260\,\AA\ long
spectral window centered on the H$\beta$ line and the polarimeter optics have been optimized to work in this wavelength range.
The spectral resolution is high enough to resolve spectral
lines in slowly rotating stars but at the same time
is low enough so that any polarization signal in the broad hydrogen line, as well as broad metallic lines
of rapidly rotating stars, can be precisely measured.

\subsection{Quarter-wave plate}

\begin{figure}[b!]
\includegraphics[angle=270,scale=.50]{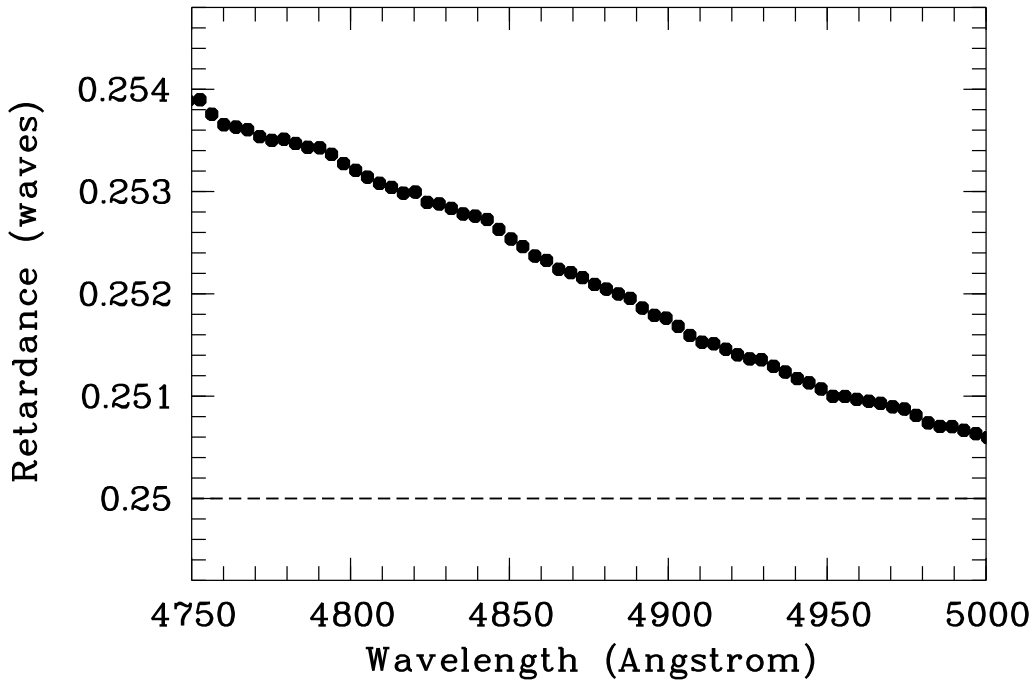}
\caption{Measured retardance of the quarter-wave plate used in {\em dimaPol}.
The nominal $\lambda/4$ retardance is shown by a dashed line.
Error bars are approximately the size of the data points.
\label{quarterwave-retardance}}
\end{figure}

The quarter-wave plate in the polarimeter module converts left and right 
circularly polarized light in the input beam into linearly polarized light with orthogonal polarization directions.
The achromatic quarter-wave plate in {\em dimaPol}
was made by Bolder Vision Optik and is a multilayer polymer laminated between two optically flat windows.
The retardance error is less than 0.004\,$\lambda$ over the wavelength
range where the polarimeter performance is optimized
(Figure~\ref{quarterwave-retardance}).
A polymer plate was selected because of its large acceptance angle,
a requirement for use in the diverging input beam from the telescope.

\subsection{Half-wave plate}

The quarter-wave plate is followed by a half-wave plate, a commercial FLC made by Displaytech Inc.\ (model LV2500-OEM).
This is an electrically switchable retardation plate with a fixed
retardation of approximately 180$^{\circ}$ or $\lambda/2$ but the 
orientation of the optical axis can be switched between two distinct angles by
changing the voltage applied across the liquid crystal.
A commercial driver DR95 made by Displaytech Inc.\ is used to supply
and switch the voltage. Switching commands are issued by the CCD controller.
For FLC wave plates this switching can be very fast.
In solar astronomy FLC wave plates have been used in a number of polarimeters,
for example,
the Zurich Imaging Polarimeter \citep[ZIMPOL,][]{gisler04},
the SOLIS Vector-Spectromagnetograph \citep{keller03},
LPSP \& TIP at the solar telescopes on the Canary Islands \citep{martinez99},
and the spectrograph of the Pic du Midi Turret Dome \citep{malherbe04}.
Modulation rates of up to 1\,kHz are achieved with the ZIMPOL's specially
designed CCD and are also quite high in the other instruments mentioned above.

A FLC plate was chosen mainly for two reasons: low cost and the ability
to switch fast. We wanted to explore the possibility of improving
the accuracy of spectropolarimetric measurements using fast modulation.
Pockels cells and
piezoelastic modulators, both capable of switching at a rate of many kHz,
were also examined
but we found these devices to be too costly for our very limited budget.
A piezoelastic modulator does not work at low frequencies and therefore
would require a special masked detector which we do not have access to.
Mechanical rotation would not allow us to explore the potential advantages of fast modulation
and is also more expensive than a FLC plate.
Another type of liquid crystal, liquid crystal variable retarders or LCVRs,
could have been employed in a fast switching polarimeter.   Such devices 
are used, for example, in the HiVIS spectropolarimeter on the 3.67-m
AEOS telescope \citep{harrington10}.
It should be noted that LCVR plates are almost two orders of magnitude
slower than FLC plates.
After considering several different options we decided that
an FLC plate was the most cost effective option allowing us
to explore fast modulation.

\subsection{Modulation}

The role of the half-wave plate is to switch the orthogonal states of linear polarization
produced by the preceding quarter-wave plate.
When the optical axis of the half-wave plate is aligned with
the polarization plane the linearly polarized beam passes through
the half-wave plate unchanged; when the optical axis
of the half-wave plate is at an angle of 45$^{\circ}$ to the polarization plane it
rotates the linearly polarized light 90$^{\circ}$.
As a result, the half-wave plate effectively switches the two orthogonal polarizations
by rotating both polarizations 90$^{\circ}$.

Because of the beam displacer that follows the half-wave plate (see below),  switching 
the polarization states also results in the polarimeter's two output beams exchanging places.
In {\em dimaPol} the modulation of the beams is synchronized with the CCD detector.
As discussed above, the switching needs to be relatively fast in order to eliminate instrumental effects
that can affect accurate polarization measurements.
Simply reading out the detector frequently after short exposures is not desirable since this
would significantly increase overheads because of the CCD readout time and likely also quickly 
wear out the instrument's shutter.
The SITe detector is, however, a three-phase CCD that allows bi-directional charge transfer. 
This permits rapid shuffling of accumulated charge on the CCD back and forth between rows many times before 
the detector is read out.
Such parallel transfers are almost three orders of magnitude faster than the read out time for the detector.
Rather than reading out the CCD frequently we instead shuffle the charge on the CCD up and down
with every switch of the half-wave plate.
This charge shuffling technique for measuring polarization signals
was first implemented by \citet{mclean81} in the Imaging SpectroPolarimeter
at the Royal Observatory, Edinburgh.
The shuffle distance in {\em dimaPol} is chosen to be the same
as the final separation of the two output spectra,
or about 40\,pixels in a direction perpendicular to the spectra.
During the charge shuffling the orthogonal polarization states never cross each other's path.
The CCD shutter remains open for the duration of an exposure for multiple switch cycles
and is kept closed during CCD readout.

The switching time of the FLC half-wave plate at room temperature is about 70\,$\mu$s, and
while it increases to about 120\,$\mu$s at +10\,C, switching at kHz rates is possible even 
at lower temperatures.
This is significantly faster than the approximately 2\,ms that
is needed to move the charge 40\,rows on the CCD so there is some `dead time' while charge is
being shuffled.
Since the wave plate switching time is only a small fraction of the shuffle time the switching is done in 
the middle of the charge shuffling to minimize the crosstalk between the two polarizations.

\subsection{Retardance and the switching angle of the half-wave plate}

\begin{figure}[t]
\includegraphics[angle=270,scale=.50]{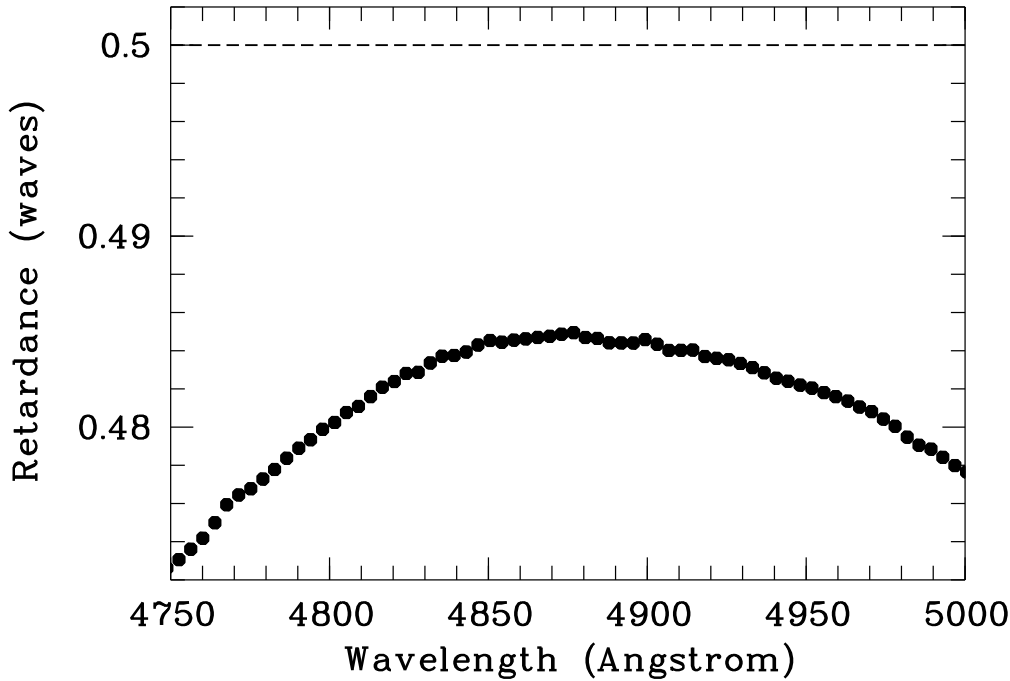}
\caption{Measured retardance of the FLC half wave plate used in {\em dimaPol}.
The dashed line is the plate's nominal $\lambda/2$ retardance.
\label{flc-retardance}}
\end{figure}

An obvious advantage of using an electrically switchable wave plate is that
there are no moving parts in the polarimeter.
This provides better stability, reliability, and significantly reduces both construction and maintenance costs.
FLC wave plates are also true zero order retarders and they have a large field of view.
The acceptance angle of our FLC wave plate is 20$^{\circ}$.
This permits its use in the $f/18$ beam from the telescope without introducing any noticeable error in wave plate retardance.
One disadvantage of FLC wave plates is the fact that they are chromatic;
retardance of the DAO FLC wave plate differs from
$\lambda/2$ by $0.015\,\lambda$ in the middle of the spectropolarimeter's spectral range and
increases to $0.027\,\lambda$ and $0.022\,\lambda$ at
the blue and red ends respectively (Figure~\ref{flc-retardance}).

The data shown in Figure~\ref{quarterwave-retardance} and
Figure~\ref{flc-retardance} were obtained in our optical laboratory.
The plate to be tested was placed first between two open (the axes parallel)
and then two closed (the axes orthogonal) BVO Vis22 linear polarizers 
from Bolder Vision Optik.
The quarter-wave plate was oriented at 45$^{\circ}$
with respect to the first fixed polarizer.
The change in the beam intensity was measured with an Ocean Optics
USB2000 fiber optic spectrometer and the wave plate retardance was calculated using the equation
$$\delta (\rm{waves}) = \frac{1}{2\pi} \arccos\left(\frac{I_{||} - I_{+}}{I_{||} + I_{+}}\right)$$
where $I_{||}$ and $I_{+}$ are the beam intensities with the polarizers
open and closed respectively.
The retardance of the FLC plate was measured in the same way.
All of the measurements were carried out at room temperature.

\begin{figure}[t]
\includegraphics[angle=270,scale=.50]{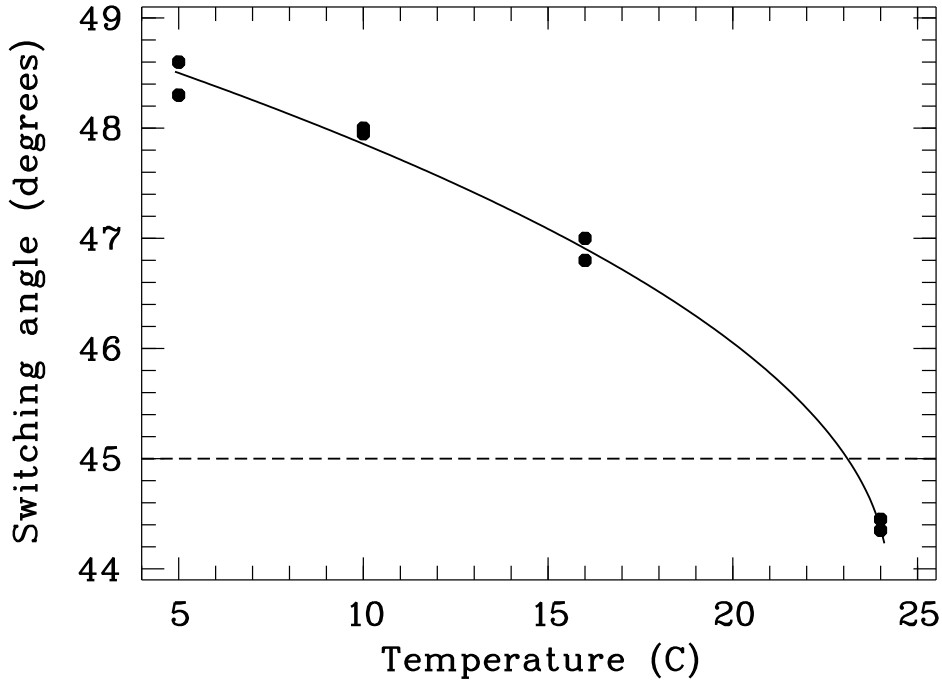}
\caption{The FLC switching angle as a function of temperature.
The data points are measurements performed in a cold chamber
at the HIA for one of our spare wave plates. 
The nominal switching angle of the wave plate is shown by the dashed line.
The solid line is a fit to the data of the form $c_1 + c_2\sqrt{(T_0 - T)}$,
where $T$ is the plate temperature and $c_1$, $c_2$, and $T_0$ are constants.
Such a temperature dependence is typical of that observed in FLC wave plates \citep{blinov94}.
\label{flc-switchangle-temp}}
\end{figure}

While the FLC plate retardance is not temperature dependent
the switching angle is.
To examine this effect we carried out an experiment in the cold chamber at the HIA where the temperature
of the wave plate could be controlled.
The wave plate was placed at 45$^{\circ}$ between two open polarizers.
The second polarizer was rotated from its ``open'' position until
a minimum in the beam intensity was reached.  The deviation angle of the second
polarizer from the ``open'' position is then  $2\times$ the departure of the wave plate
switching angle from 45$^{\circ}$.
One of our spare plates was used in this test and the measurements were performed at 4600\,\AA\
where the retardance of the plate is closest to ideal.
Figure~\ref{flc-switchangle-temp} shows the results.  
 We find that the switching angle
is exactly 45$^{\circ}$ at $+23 \pm 1$\,C, a typical summer temperature
at the DAO.
In winter Victoria temperatures drop to +5\,C and the switching
angle at this temperature is approximately $48.5^{\circ}$.
Our modelling shows that this change in the switching angle
can introduce crosstalk from linear to circular polarization
of up to 13\%. The crosstalk should be significantly less in summer.
Crosstalk from linear to circular polarization is not uncommon in
other polarimeters
but does not pose a problem when measuring circular polarization
in upper main sequence magnetic stars because their intrinsic linear polarization
is at least an order of magnitude lower than their circular polarization.
Thermal stabilization of the FLC could solve the crosstalk issue 
but has not been implemented for {\em dimaPol} because of lack
of space and the limited impact such crosstalk has on the science we anticipate performing
with the instrument.

\subsection{Crosstalk}

\begin{figure}[t]
\includegraphics[angle=270,scale=.25]{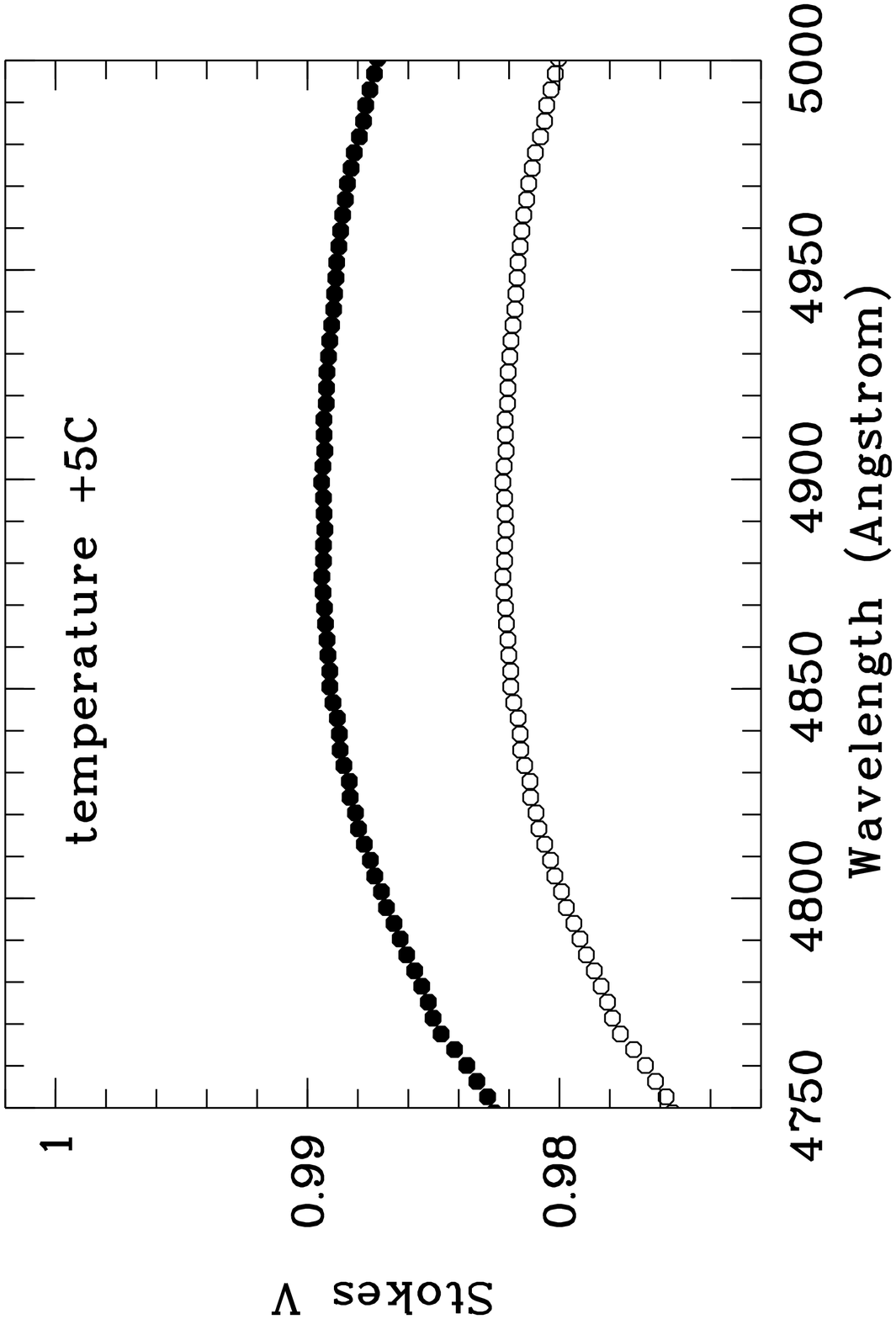}
\includegraphics[angle=270,scale=.25]{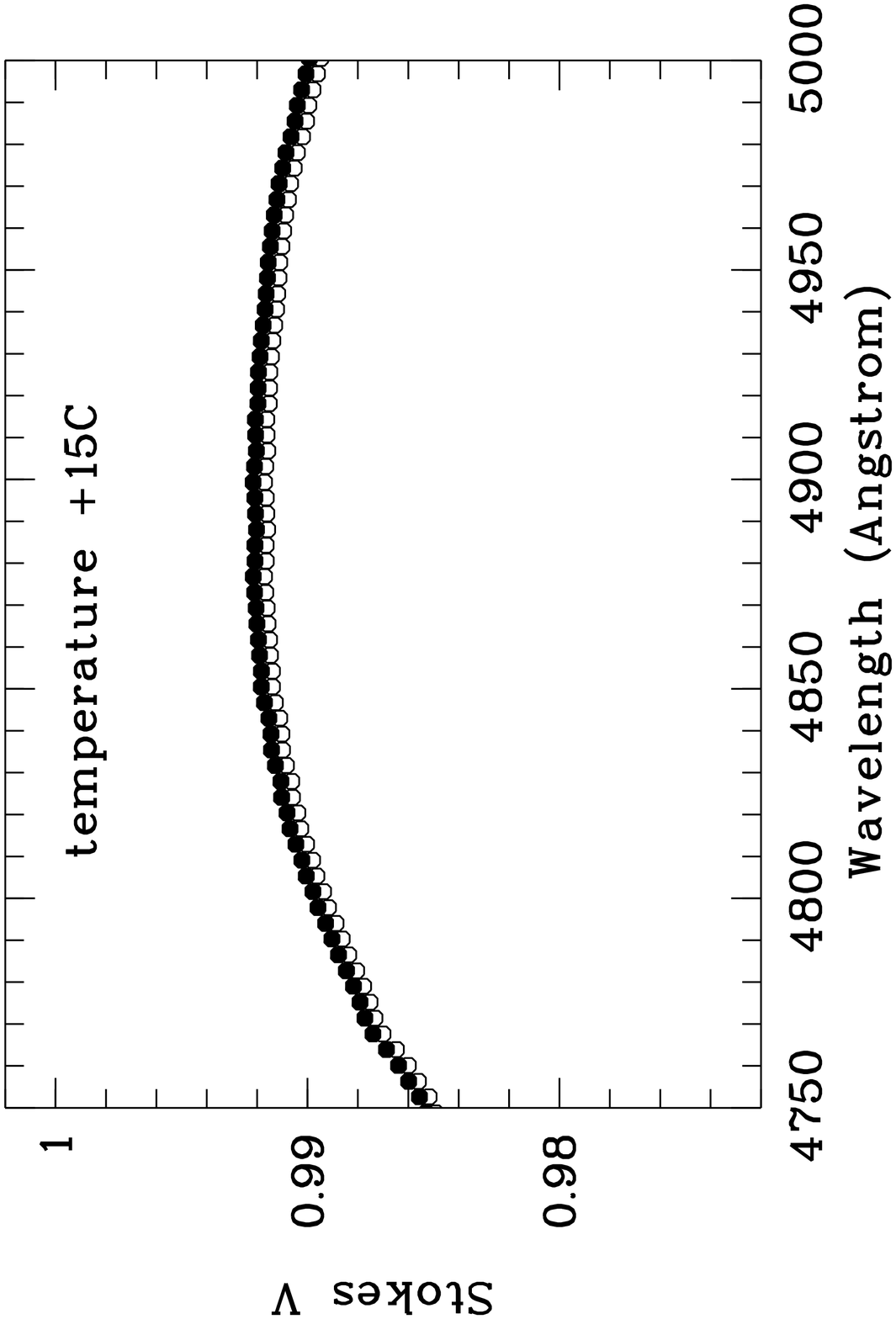}
\caption{Depolarization 
due to crosstalk between the two channels of  {\em dimaPol}  (open circles).
100\% input circular polarization is assumed (Stokes V equal to 1).
The crosstalk can be significantly reduced by rotating the FLC plate
by 2$^{\circ}$ (filled circles).
\label{crosstalk}}
\end{figure}

Departures of wave plate parameters from the ideal creates crosstalk between
the two polarimeter channels.  Some of the light from the ordinary beam
leaks into the extraordinary beam and vice versa.
As a result one will see a change in the measured Stokes V.
We used the data in Figures~\ref{quarterwave-retardance} through~\ref{flc-switchangle-temp}
to estimate how much the Stokes V measurement is affected by such crosstalk.
We obtained estimates for two different temperatures:
+5\,C and +15\,C.
We assumed that the instrument is fed with 100\% circularly
polarized light (i.e., the Stokes V is equal to 1) and the results are shown in Figure~\ref{crosstalk}.
The departure of the Stokes V from 1 tells how much crosstalk there
is in the system.
As expected, the amount of crosstalk is higher at the lower temperature
mostly due to the increase in the FLC switching angle from 47$^{\circ}$
at +15\,C to 48.5$^{\circ}$ at +5\,C.
This crosstalk can be significantly reduced by rotating the FLC
plate by 2$^{\circ}$ around the beam and in the direction opposite
to the wave plate axis tilt.  The amount of crosstalk is reduced by a factor
of two at the lower temperature and decreases insignificantly
at the higher temperature.
As a consequence, in the case of a simple Zeeman triplet and a magnetic field aligned
parallel to the line of sight {\em dimaPol} will underestimate
the longitudinal field component by about 0.5 to 1\%.
However, longitudinal magnetic fields are very rarely measured with an accuracy
sufficient to detect such a small discrepancy \citep[for example,][]{bychkov05}.

The difference in transmittance of the FLC wave plate along the fast and slow axes 
(up to a few percent), together with different transmittance for ordinary and extraordinary beams
inside of the spectrograph, results in spurious continuum polarization.
The amplitude of this signal reaches 1\% at some wavelengths
but since this unwanted signal is much broader
than any spectral line in the region, including H$\beta$, and is
a smooth function of wavelength it can be easily filtered out during data reduction as discussed below.

\subsection{Beam displacer}

A 20\,mm long calcite beam displacer separates the orthogonal linear polarizations 
into two parallel beams.
The ordinary beam with the plane of polarization along the dispersion
travels straight through the displacer while the extraordinary beam emerges 
from the beam displacer with a constant displacement of about 2\,mm in the direction
perpendicular to the dispersion.
Because of this, after passing through the spectrograph the two beams
arrive at two different locations on the detector where
both orthogonal states of polarization can be measured simultaneously.
At the detector the spectra are separated by $\approx 0.6$\,mm or $\approx 40$\,pixels.
Again, it is important to note that the two beams never move on the detector and that 
there are no moving parts in the polarimetric module.
The two beams always illuminate the same rows of the detector.

\begin{figure}[t!]
\includegraphics[angle=0,scale=.48]{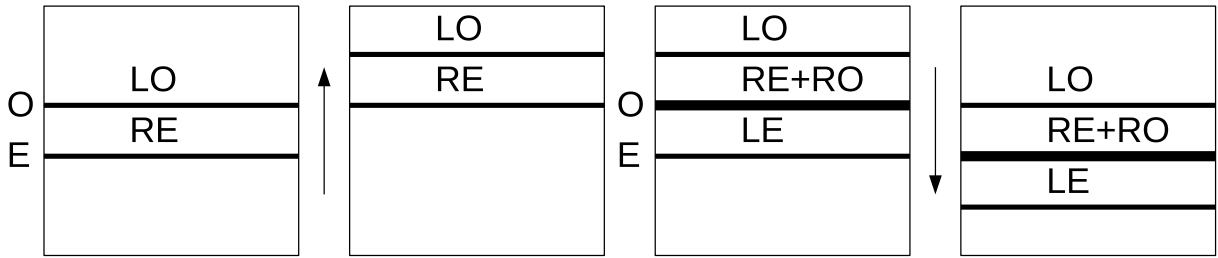}
\caption{A schematic representation of one switch cycle of {\em dimaPol}.
From left to right: expose, move the accumulated charge up, switch
the FLC wave plate and expose again, move the charge back down and switch
the wave plate in preparation for the start of the next switch cycle.
As a result, three spectra form on the CCD (see text).
\label{shuffle}}
\end{figure}

With no half-wave plate switching and no CCD charge shuffling the observation of a star
with {\em dimaPol} would obviously result in two spectra being imaged on the CCD,
but with very poor polarimetric precision because of instrumental effects mentioned above.
The implementation of charge shuffling results in three spectra being produced as follows
(and illustrated in Figure~\ref{shuffle}).
At the start of an exposure the left circularly polarized light traveling through
the polarimetric module becomes the ordinary beam while
the right circularly polarized light becomes the extraordinary beam.
As a result two spectra, LO and RE, appear on the CCD.
Then, when the FLC wave plate switches (and ignoring for now the fact that the charge shuffling
takes considerably longer than the wave plate switch) the
charge in the LO and RE spectra is shifted upward so that
RE is brought up to where the ordinary beam is projected on the CCD and
LO is  placed 40 rows higher, outside of the light collecting area.
But at the same time, because of the wave plate switch the right circularly polarized light 
becomes the ordinary beam (RO) and so it is projected right onto RE.
RO and RE then mix together to form a single right circularly polarized spectrum.
The left circularly polarized light becomes the extraordinary beam
after the wave plate switch and because of the charge shuffle this creates a third spectrum (LE).
Now there are three spectra on the CCD: LO, RO+RE, and LE.
When the modulator switches again all three spectra are shuffled 40\,rows back down
and the cycle repeats.
How these three spectra are processed and manipulated to measure a stellar magnetic field is discussed in the next section.

Since only a single beam displacer is used in {\em dimaPol},
the ordinary and extraordinary beams travel slightly different optical path lengths and therefore
have different foci. 
ZEMAX models show that the difference in the best focus position
is very small, approximately 20\,$\mu$m.
When the spectrograph is focused at an intermediate position (when both images are slightly
out of focus) the degradation in image quality amounts to less than 2\,$\mu$m or
roughly 0.1\,pixel.
This is less than our typical focus uncertainty and can, therefore,  be neglected.
The difference in optical path between the two beams is also wavelength dependent and
as a result the size of the extraordinary image along
the dispersion is about 15\% larger at the blue end of the detector.
There is also a small difference in dispersion between the two beams.
Despite these effects, when the ordinary and extraordinary beams are combined
(see the next section) the spectral resolution is reduced by less than 5\% and so we
felt there was little need to use a much more costly Savart plate instead of a single
calcite crystal.
Savart plates are comprised of two plates
of birefringent material rotated 90$^{\circ}$ with respect to each other.
The extraordinary beam in the first plate then becomes an ordinary beam
in the second plate and vice versa and, as a result, both beams have nearly identical
optical paths and focus positions.

After the beam displacer a fast mechanical Uniblitz shutter is installed.
The shutter has opening and closing times of 18\,ms and 12\,ms respectively.

All optical surfaces in the polarimetric module are antireflection coated.
The transmission of the polarimeter module is therefore quite high and varies
smoothly between 74 and 80\% within its operational wavelength range.

\section{Fast Switching and Observing Strategy}

We have already noted that switching a spectropolarimeter quickly helps to reduce
the impact of spurious signals that can arise from changes in the spectrograph aperture
illumination due to seeing and transparency variations, guiding errors,
and other telescope instabilities.
Some of these changes in illumination can be averaged out during long exposures
but trends that manifest themselves only on time scales of hours are not compensated for in this way.  
It therefore seems that a better observing strategy for effectively eliminating the effects of changes in aperture illumination is to take multiple short exposures,  switch the polarimeter module after each, and then combine the resulting sequence of exposures.
When a short switching time is used it is much more efficient to do
the averaging on-chip by moving the accumulated charge back and forth rather than after many CCD readouts
and this is what we do with {\em dimaPol}.

But the question arises as to how many times the polarimeter module needs to be switched in order to provide the most
effective averaging and hence elimination of spurious signals.
We have performed a series of on-sky experiments in an attempt to answer this question.
In these tests we kept the total integration time on the target constant
but varied the shuffle time or the number of shuffles per exposure.
We define the shuffle time as the amount of time
used to expose briefly, switch the polarimetric FLC half-wave plate, and move
the charge on the detector. 
The number of shuffles is simply the total exposure time divided by the shuffle time.
Only an even numbers of shuffles are used to make sure that the polarized spectra obtained 
for successive targets always appear in the same place on the CCD. 

A cycle of two shuffles therefore consists of the following sequence: expose, switch the wave plate, move the charge, expose, switch the wave plate  to its original state and move the charge back.
As noted earlier, it takes approximately 2\,ms to move the charge 40 rows on the CCD while 
the switching of the wave plate is considerably faster and is performed at the middle of the shuffling phase. 
The total dead time per shuffle is, therefore, approximately 2\,ms and this sets a lower limit to the
possible shuffle time to ensure an efficient duty cycle. 
We refer to this as `dead time' since during this period some of the light on the detector gets spread
in between the spectra and this signal is removed, together with the rest of the background signal, during processing 
(described below).

For our switching tests we observed Vega which, along with its brightness, 
offers the advantage of having narrow spectral lines and hence makes it an easy 
target to obtain high quality magnetic field measurements in a very short period of time.
The  exposure time for individual observations in a given observing sequence
 was kept constant at 6\,s to provide approximately 
the same signal per exposure, while the shuffle time 
from observation to observation was varied between 0.06 and 3\,s 
(corresponding to a range of 2 to 100 shuffles per exposure).
CCD charge trapping sites begin to manifest themselves when a few hundred
shuffles or more are executed \citep{bland95}.
These traps create spikes in the extracted spectra, polluting the polarization
signal. In order to avoid this effect we limited ourselves to 100 shuffles.
On three different nights we obtained a total of 21 magnetic field measurements derived from the H$\beta$ line.
Each measurement was based on the average of 30 individual 6\,s exposures.
We did two things to minimize the effect of changing weather conditions 
and flux variations on our results:  first, the shuffle time was changed
quasi-randomly from measurement to measurement but within the 0.06 to
3\,s range, and second the longitudinal field error bars were normalized
by the flux as $\sigma(B_l)_{norm} = \sigma(B_l) \sqrt{flux / flux_{max}}$.

\begin{figure}[t]
\includegraphics[angle=270,scale=.50]{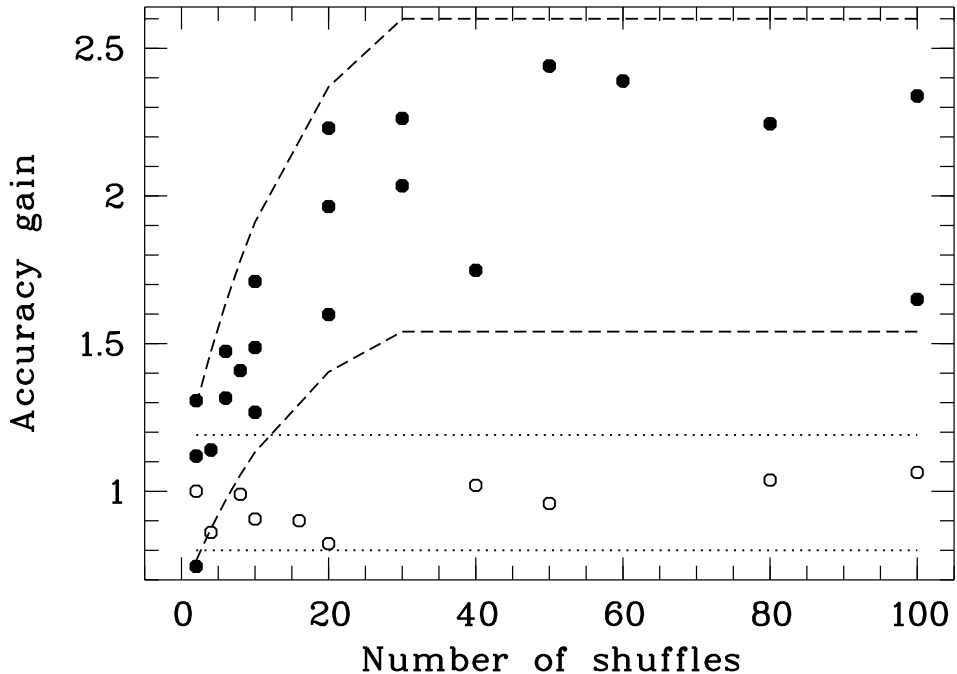}
\caption{Measured variations in the accuracy of longitudinal magnetic field
measurements of Vega (filled circles) and scattered solar light (open circles)
as a function of the number of wave plate switches and charge shuffles
executed by {\em dimaPol}.
The gain in accuracy is measured in comparison to the average estimated
magnetic field error when only two shuffles are carried out.
95\% confidence intervals of the standard deviation are
also shown for Vega (dashed lines) and for scattered solar light
(dotted lines).
\label{accuracy_vs_nshuffles}}
\end{figure}

Figure~\ref{accuracy_vs_nshuffles} shows how the accuracy of magnetic field measurements
obtained with {\em dimaPol} vary as a function of the number of shuffles.
The magnetic  error bar is largest when only two wave plate switches and charge shuffles are carried out.
The gain in accuracy has therefore been measured by normalizing the estimated error in the 
longitudinal magnetic field for an arbitrary even number of shuffles to the mean value found
for observations with just two shuffles.
The magnetic field error bars get smaller as the number of shuffles increases until more or less a plateau is reached at about 30 shuffles.
The accuracy is improved by a factor of 2 when 30 or more switches
are executed.
We also carried out observations of scattered solar light.
The telescope was pointed at the sky near the zenith during the day
and series of 50 short 4\,s exposures were obtained for several different
shuffle speeds.
Unlike the stellar measurements, these scattered solar light observations
are not affected by slit losses and can therefore be used
to test for systematics within the spectropolarimeter itself, while excluding
other telescope instabilities.
The results are presented in Figure~\ref{accuracy_vs_nshuffles} along
with the results for Vega.
95\% confidence intervals of the standard deviation are
also shown for both Vega and scattered solar light.
For Vega a third degree polynomial approximation of points was used
for $N \le 30$ and the average accuracy gain of 2.08 was assumed for $N > 30$.
The points are within the confidence intervals for both data sets.
There is no change in the accuracy gain seen for
the scattered solar light data which means that most likely slit losses are
responsible for the accuracy degradation in Vega at low switching speeds.
Switching more than 30 times eliminates the effect these losses have
on the longitudinal magnetic field measurements.

As a result of these tests, we have adopted a standard observing strategy for
use with {\em dimaPol}.
To make a single magnetic field measurement we obtain a series of typically 10 to 16 short 
exposures (each, for example, 5 or 10\,min long for a sixth or seventh magnitude star) of the target  
with 60 wave plate switches and charge 
shuffles performed during each exposure.   
This keeps the switching time shorter
than the typical time of instrumental response changes, and also allows an estimate of the
magnetic field uncertainty to be made from the scatter of measurements obtained from the individual exposures.

Because exposure times are short, we also bin every two pixels of the SITe CCD along the
direction perpendicular to the dispersion to speed up the readout of the CCD and 
reduce the impact of readout noise.
Since typical seeing at the DAO is around $2-3$\,arcsec,
we use the 2\,arcsec entrance aperture
for all observations with the polarimeter.
We also acquire 60\,s iron-argon comparison lamp exposures before the start and at the conclusion of the
observing sequence for each star.  
The polarization module switches at 1\,Hz for the arc exposures so that we have the usual 60 
switches and shuffles for the arc observations that we use for our science exposures.

\section{Data Reduction}

We have developed a collection of MIDAS scripts that enable near-real-time processing of the data
produced with {\em dimaPol}.
This software provides a measurement of the longitudinal magnetic field and its
associated uncertainty within a few minutes of the acquisition of the last spectrum
in an observing sequence for a star.
For every individual exposure we perform bias subtraction, cosmic ray
and background removal, and then extract the spectra.
The bias level in each image is determined by measuring just the CCD overscan area
and subtracting this from all of the other image pixels. 
The remaining background is removed in the next step.

As has been mentioned earlier, the CCD shutter remains open for many wave plate switch and
charge shuffle cycles.
Because of this, even though the parallel charge transfer takes very little time, some light from the 
source does get spread out between the three recorded spectra during the shuffling process and this
creates an additional background source on the detector. 
This extraneous light needs to be removed before the polarized spectra can be extracted.
The background flux has a trapezoid like shape when viewed along the CCD columns.
We divide each CCD image into three sections, one for each of the three spectra,
and treat the background in each section separately.
In each of the two outer sections we fit second order polynomials to the background flux measured on 
either side of each spectrum; a simple first order polynomial is used to average the background flux
on either side of the spectrum in the middle section.
These three independent polynomials are fit to every CCD column 
and the resulting background fit for each column is then subtracted from the 
corresponding column in the original image.

The next processing step is the extraction of the polarized spectra.
The three spectra (denoted LO, RO+RE, and LE above) are extracted separately.
During the spectrograph, CCD and polarimeter configuration carried out in advance of
each polarimeter run,
considerable care is taken to ensure that the acquired spectra run parallel to the CCD rows.
The spectrum inclination and curvature are then typically less than 5\,$\mu$m over the approximately
26\,mm length of CCD.
This makes the spectral extraction process quite simple:
CCD rows inside of a predefined aperture are simply co-added.
We have determined that the best results are obtained when five CCD rows are coadded.
Since the CCD is binned by a factor of two perpendicular to the dispersion this corresponds to ten
unbinned rows on the detector.
Such an aperture gives the best signal-to-noise ratio when a 2\,arcsec spectrograph circular
entrance is used.
The size of the aperture is fixed and the same size is used to extract all three spectra.
The three apertures are placed 40\,pixels from each other (corresponding exactly to the charge 
shuffle distance) to make sure that we always extract the same pixels for the RO+RE, LO, and LE spectra.
The two polarizations are exposed on exactly the same set of physical pixels even
though the charge during shuffling may be stored in different places.
After extraction the LO and LE spectra are combined into a single spectrum, LO+LE.

As alluded to earlier, the FLC wave plate and other optical elements in the spectrograph
introduce an extremely broadband spurious continuum polarization that can produce a weak 
circular polarization signal, even from an unpolarized light source.
The effect is wavelength dependent and for an unpolarized light source leads to
an LO+LE spectrum that is $\approx 1\%$  brighter than 
the RO+RE spectrum at both ends while the two spectra have almost identical 
intensities in the middle.
Even though this broadband signal changes very little from
exposure to exposure and from object to object, it can affect precise line shift measurements required
for magnetic field observations since the small wavelength-dependent difference in brightness between the two polarizations results in an artificial line shift.
It is also dependent on the size of the spectral window used for measuring pixel shifts and 
discussed below: the broader the spectral window the larger the introduced pixel shift.
If not taken into account during the data processing, this instrumental shift can be as large as a 
few hundredths of a pixel which, as will be seen,  can lead to a spurious, and obviously significant, 
signal on the order of a few 100\,G. 

In order to remove this effect we apply a chromatic correction to our data.
A third degree polynomial is sufficient to fit the variation in $I_{\rm LO+LE} / I_{\rm RO+RE}$ 
along the direction of the dispersion.
The unnormalized RO+RE spectrum is then multiplied by this fit to
eliminate the artificial brightness difference
between the two polarizations, and removes the spurious line shift as well.
Note that the chromatic correction is only derived from and applied to the final, combined
LO+LE and RO+RE spectra for an entire observation and not to individual exposures.
This minimizes the possible impact of low signal-to-noise levels in individual observations
on the polynomial fit for the correction.
We feel that normalizing $I_{\rm LO+LE} / I_{\rm RO+RE}$ is far preferable than trying to
normalize the intensities of the LO+LE and RO+RE spectra independently. 
The spurious continuum polarization as it manifests itself in $I_{\rm LO+LE} / I_{\rm RO+RE}$
can be fit very well with a low order polynomial. 
On the other hand the continuum in
the intensity spectrum is considerably more complex in shape and is more
difficult to reconstruct, especially when broad spectral lines are present and when the
normalization has to be carried out in a completely automated fashion for stars with a
wide range of spectral types during data acquisition.

Finally, the magnetic shift observed in a single spectral line due to the Zeeman effect
is then measured by performing a Fourier 
cross-correlation of the final combined and corrected LO+LE and RO+RE spectra in a spectral window
centered on the spectral line of interest.
Several different methods of measuring the magnetic shift
in a single line have been used by different authors.
These include the center of gravity method \citep{rees79} and its
implementations by \citet{mathys88} and \citet{donati97},
the so-called photopolarimetric method \citep{bray64} adapted by
\citet{borra73} for use with a two-channel Balmer-line photoelectric
Pockels cell polarimeter and by \citet{bagnulo02} for use with
the FORS1 spectropolarimeter,
and the Fourier cross-correlation technique \citep{monin02}.
We have adopted the Fourier cross-correlation method for use
with {\em dimaPol} because of the technique's relative
insensitivity to continuum and line shapes which permits its use
with unnormalized spectra and without prior knowledge of the intrinsic
profile of the line in the absence of a magnetic field.

The magnetic shift is translated into a longitudinal magnetic field, $B_l$ (G),
according to the line's magnetic sensitivity or the Land\'{e} factor and the relation
\citep[for example,][]{landstreet92}
$$B_l = \Delta X \times 0.15 / (2 \times 4.67 \times 10^{-13} \lambda^2 g),$$
where $\Delta X$ is the observed shift between the spectra in pixels, $\lambda$ is the wavelength 
of the line center (\AA) and $g$ is the Land\'{e} factor for the line.
For hydrogen lines the Land\'{e} factor is 1 \citep{casini94} while for
metallic lines the Land\'{e} factors are taken from the VALD database \citep{kupka00}.
The measured shift in pixels is multiplied by the average dispersion of
0.15\,\AA\,pixel$^{-1}$.
The change in dispersion that arises along the spectrum as a result of the calcite beam displacer
is small and can be neglected.
For H$\beta$ observations with {\em dimaPol} this equation is transformed to the simple relation 
$B_l = 6.8 \times \Delta X$\,kG.
In other words, a one pixel shift between the RO+RE and LO+LE spectra corresponds to a longitudinal
magnetic field strength of 6.8\,kG.

More than ten individual observations, and hence magnetic measurements, are 
usually obtained in succession for a typical $V=7$ target and many more are often acquired
for observations of very bright stars (usually known magnetic `standards' or `nulls' observed during
each observing run to monitor performance of the instrument).
The multiple independent longitudinal field measurements are then averaged
and an error bar can then be derived from the scatter of these individual values, after weighting by the
flux as discussed in the previous section.

\begin{figure}[t]
\includegraphics[angle=270,scale=.50]{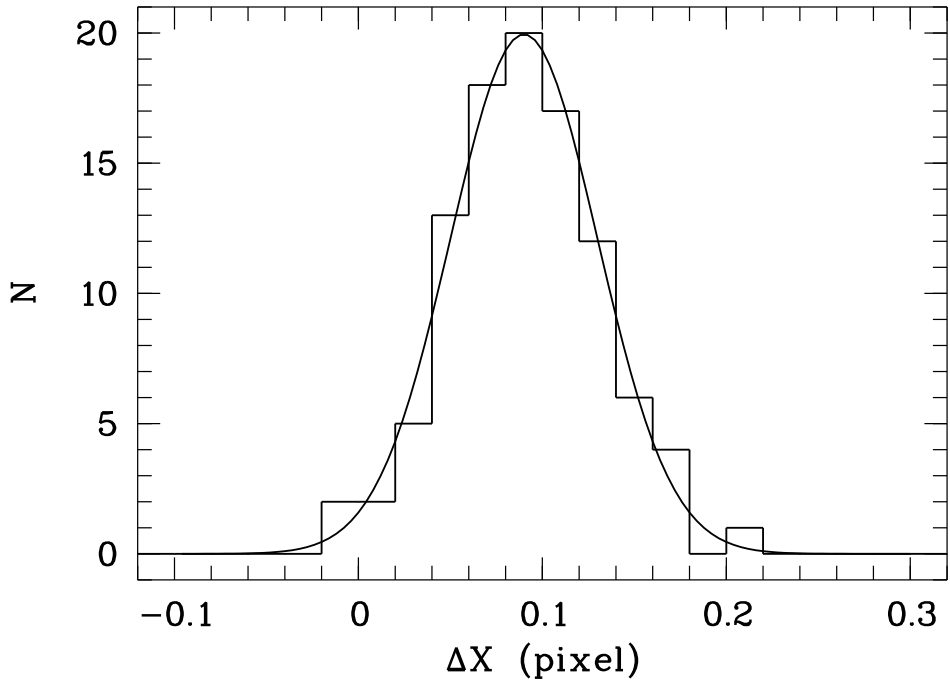}
\caption{An illustration of the apparent normal distribution of individual magnetic field observations
obtained with {\em dimaPol}.  The histogram shows the distribution of 100 individual 
measured shifts (in pixels) between left- and right-circularly polarized spectra obtained during a
single observing sequence of the bright, magnetic Ap star $\alpha^2$\,CVn.
The superimposed curve shows an expected normal distribution with a mean of 0.09\,pixels 
and variance of $(0.04$\,pixels$)^2$.  
A Kolmogorov-Smirnov test confirms that the observed pixel shifts are indistinguishable from a 
normal distribution.
\label{norm-distribution}}
\end{figure}

Figure~\ref{norm-distribution} shows a histogram of 100 measurements
of the shift between the RO+RE and LO+LE spectra for the bright, sharp-lined 
Ap star HD\,112413 ($\alpha^2$\,CVn) obtained in succession
over almost one hour on a single night.
The mean shift between spectra is $\approx 0.09$\,pixels corresponding to a field strength of 
$\approx 600$\,G and the sample standard deviation is $\approx 0.04$\,pixels
or $\approx 270$\,G.
The measurements are apparently distributed normally as a curve representing a normal 
distribution with a mean of 0.09\,pixels and variance of $(0.04$\,pixels$)^2$ fits the histogram 
of observations very well.
A Kolmogorov-Smirnov test confirms that the measurements are very
likely normally distributed: the P-Value of 0.98 indicates
that the histogram points and the normal distribution differ insignificantly.
When the points are normally distributed the standard error of the mean
is proportional to $N^{-1/2}$ where $N$ is the number of observations. 
Therefore, it seems quite reasonable in the case of {\em dimaPol} data to estimate the 
magnetic field error bars by multiplying the standard deviation of the individual observations
by this same factor.
In the above example the error bar is then
$0.04 \times 100^{-1/2} = 0.004$\,pixels or $\approx 27$\,G.

\begin{figure}[t]
\includegraphics[angle=270,scale=.50]{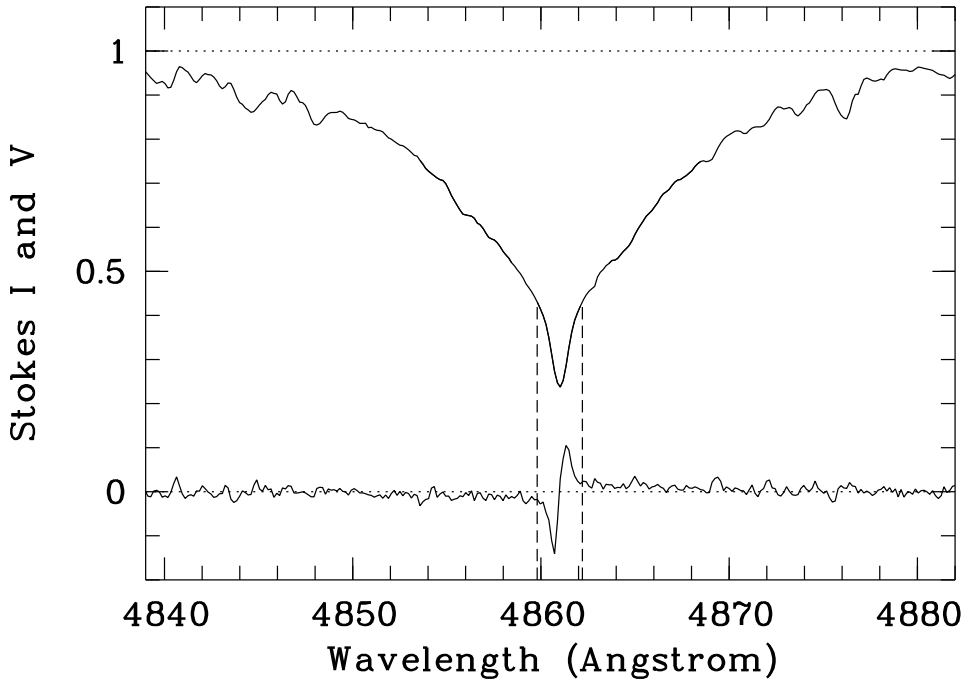}
\caption{An example of {\em dimaPol} observations of the H$\beta$
line of the prototypical magnetic Ap star $\alpha^2$\,CVn.
The Stokes $V$ signature is at the bottom and has been multiplied by a 
factor of 10 relative to the $I$ spectrum.
Uncertainties in longitudinal field measurements derived from such data are
minimized by optimizing the size of the spectral window used for the Fourier
cross-correlation used to measure the shift between the LO+LE and RO+RE
spectra by ensuring that it only includes the portion of the line profile
that gives the strongest polarization signal.
For $\alpha^2$\,CVn the spectral window is indicated by the vertical dashed
lines and is 16\,pixels wide.
For more rapidly rotating stars the window can be up to 128\,pixels wide, or about 19\,\AA.
\label{hbeta}}
\end{figure}

Experimentation has demonstrated that the error bars are minimized by optimizing the size of the
window used for the Fourier cross-correlation so that the window includes the part of the profile
that gives the strongest polarization signal (Figure~\ref{hbeta}).
The window size ranges from 16\,pixels or about 2.4\,\AA\ for narrow-line stars (such as $\alpha^2$\,CVn)
to 128\,pixels or 19\,\AA\ for broad-line stars.

A careful reader may think that we have omitted some details of our data reduction procedure but
this is not the case.
Flat fielding of the DAO spectropolarimetry is not necessary for our primary goal of measuring longitudinal magnetic fields.
As mentioned earlier, both of the left- and right-circularly polarized spectra
are exposed on exactly the same CCD pixels.
As a result, pixel-to-pixel sensitivity variations are the same
for both and do not affect the final magnetic line shift determination. 
We do, however, normally obtain sequences of high signal-to-noise flat field spectra several 
times during each observing run in case they are needed in the future, for example, for a detailed analysis of Stokes $I$ line profile variations.

The use of a Fourier cross-correlation algorithm to derive the observed longitudinal magnetic field
also simplifies processing by removing any need for continuum normalization.
Thanks to the fast switching of the polarimeter module,
variations in spectrum brightness due to changing sky conditions,
slit losses, and other instrument changes are effectively averaged
for both polarizations. This results in the shape of the continuum
being almost identical in both polarizations.
The Fourier cross-correlation algorithm that we use is insensitive to features that have 
a similar shape in both polarizations and so the algorithm can be successfully applied to 
unnormalized spectra which greatly facilitates the near-real-time processing of the 
polarimeter data while observing.

Wavelength calibration of the spectra is also not  required for our magnetic field
measurements since the Fourier 
cross-correlation of the polarized spectra is carried out in pixel space.
Wavelengths are generally only needed for line identification
and for this we use the two iron-argon comparison lamp exposures obtained before the start and
at the conclusion of each sequence of observations for a given science target.
The comparison spectra are processed in the same way as the science spectra, 
the arc lines are identified, and a second order polynomial is used to derive the dispersion relation.

In addition to providing measurements of longitudinal magnetic fields,
{\em dimaPol} also allows investigations of the circular polarization
signal within the profiles of many spectral lines, including H$\beta$.
The circular polarization signal, Stokes $V$, is calculated very simply as
the difference in brightness between the LO+LE and RO+RE spectra,
normalized by the total flux:
$$V = (I_{\rm{LO+LE}} - I_{\rm{RO+RE}}) / (I_{\rm{LO+LE}} + I_{\rm{RO+RE}}).$$
An example of a short section of a Stokes $V$ spectrum can be seen in Figure~\ref{hbeta}.

The presence of the spurious continuum polarization introduced by the FLC wave plate
and other optics and the crosstalk from linear to circular polarization makes it very difficult
to measure  circular {\em continuum} polarization with {\em dimaPol}.
However, in sources with a high degree of circular polarization and an insignificant amount of 
continuum linear polarization this can be done, albeit with somewhat limited precision.

\section{Magnetic Field Observations with {\em dimaPol}}

Details of observations of the first new magnetic star discovered 
with {\em dimaPol} have recently been presented by \cite{bohlender11}.
Along with such ongoing science programs, however, regular testing of the polarimeter is 
also performed to confirm that the instrument does not produce spurious polarization signals 
that could lead to false magnetic field detections.
To ensure that this is the case, during every scheduled polarimeter run we observe magnetic
`null' standards, bright stars that have been observed with high precision numerous times with a
variety of other polarimeters and have shown no evidence for detectable magnetic fields at a level of a few tens of G.
Seven such stars have been repeatedly observed by us over a two year period.
These include HD\,886 ($\gamma$\,Peg), HD\,36486 ($\delta$\,Ori A), HD\,76644 ($\iota$\,UMa),
HD\,97633 ($\theta$\,Leo), HD\,156164 ($\delta$\,Her), HD\,172167 (Vega), and 
HD\,210027 ($\iota$\,Peg).

\begin{figure}[b!]
\includegraphics[angle=270,scale=.50]{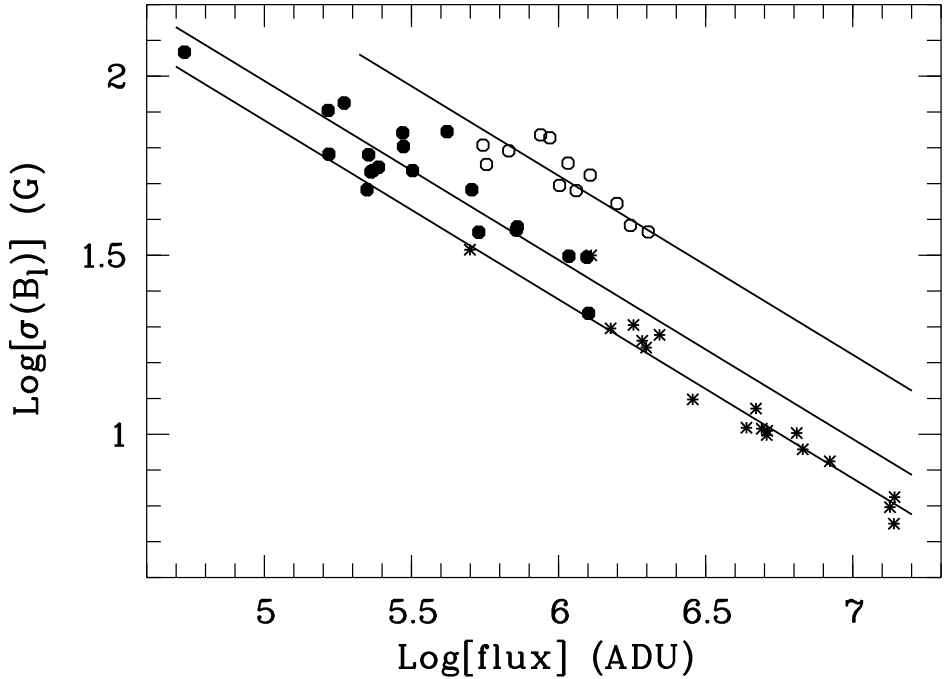}
\caption{
The uncertainty of the longitudinal magnetic field as measured with
{\em dimaPol} as a function of source flux.
Data for the Sun and two magnetic null standard stars are shown: stars - scattered solar light,
filled points - $\iota$\,Peg, open points - $\gamma$\,Peg.
The solid lines show the $\sqrt{flux}$ fit for each source.
\label{accuracy_vs_flux}}
\end{figure}

Both slowly and rapidly rotating stars are included in this collection of standards, 
with $v\sin{i}$ ranging from 3\,km\,s$^{-1}$ to 290\,km\,s$^{-1}$.
133 individual magnetic field measurements have been obtained from observations of 
the H$\beta$ line and they are all within $\pm 3 \sigma$ of 0\,G.
The average error bar, $\sigma_{B_l}$ is 78\,G and the best is 21\,G.
The null standards do not show any systematic errors at this level
and data for some of them, shown in Figure~\ref{accuracy_vs_flux},  also demonstrates that the 
measured longitudinal magnetic field error bars scale with the square root of the flux 
as expected.
Using the single H$\beta$ line, an accuracy of 20 to 40\,G is reached
in 20\,min in good weather conditions for our slow rotating ($<20$\,km/s)
third to fourth magnitude standards. 
Obviously the accuracy is lower for faster rotating stars simply because of the
decrease in sensitivity of the cross-correlation measurements as the spectral lines become broader.
For example, for HD\,76644 ($V=3.1$; $v\sin i = 150$\,km/s) we obtain
$70-80$\,G error bars in about 30\,min.
For the typical sixth to seventh magnitude stars in our science programs an accuracy
of 100 to 200\,G is reached in between one and two hours.

\begin{figure}[b!]
\includegraphics[angle=270,scale=.50]{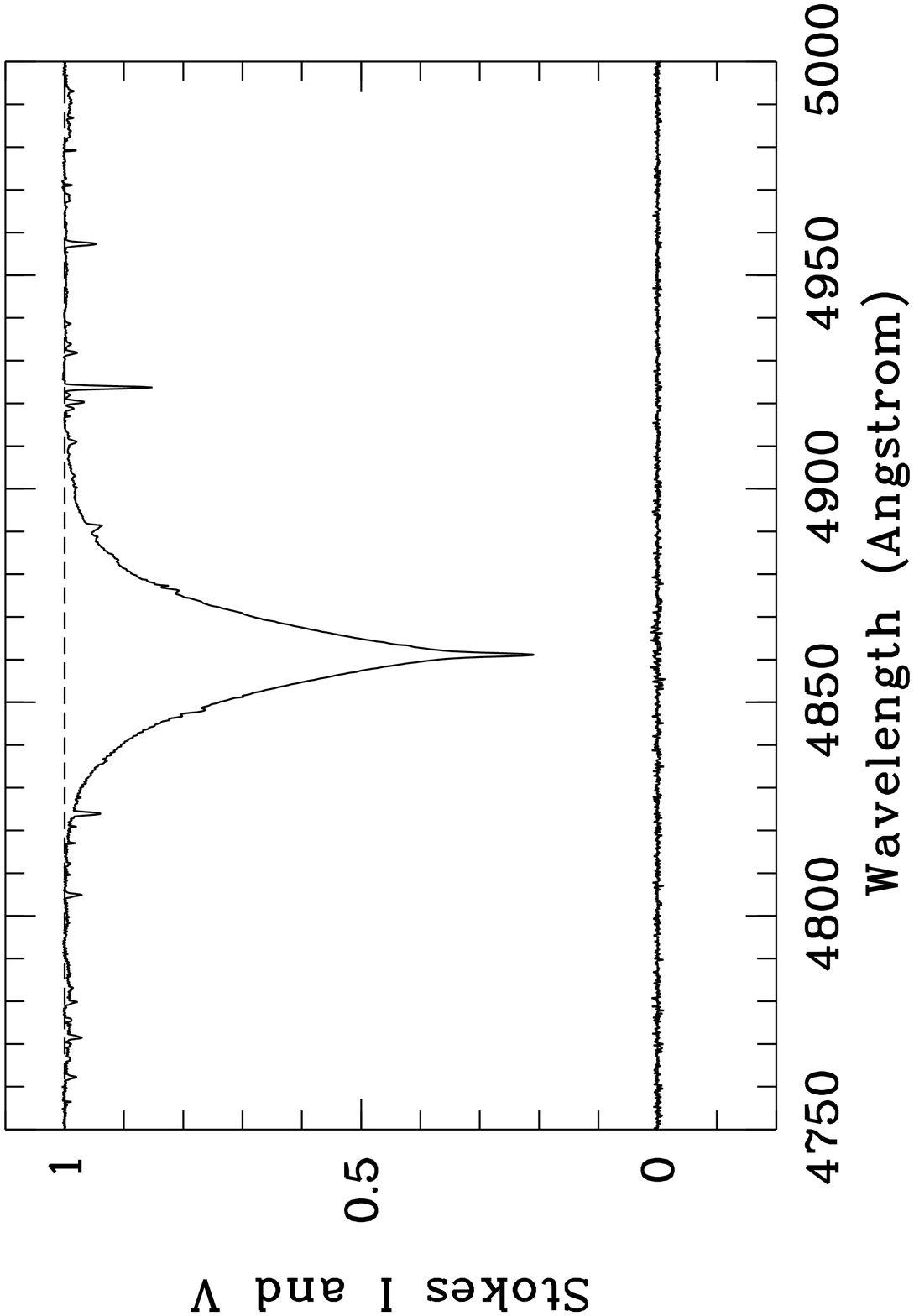}
\caption{A typical {\em dimaPol} observation of the magnetic null standard star Vega.
The Stokes $I$ spectrum is shown at the top of the figure while the lower spectrum is the
Stokes $V$ signature multiplied by a factor of 10 relative to the $I$ spectrum.
The total exposure time of the observation was 3\,min and the longitudinal
magnetic field strength derived from the H$\beta$ line is 8 $\pm$ 22\,G.
\label{vega-stokesIV}}
\end{figure}

Vega has recently been shown to  possess an extremely
weak magnetic field \citep[$<1$\,G, ][]{lignieres09,petit10} but a field of this level 
is well beyond the detection capability of {\em dimaPol} and therefore
we continue to use this convenient, bright star as a magnetic null standard.
As shown in (Figure~\ref{vega-stokesIV}), no evidence for a magnetic field is seen
in the Stokes $V$ signature of Vega as observed with {\em dimaPol}.

\begin{figure}[t]
\includegraphics[angle=270,scale=.50]{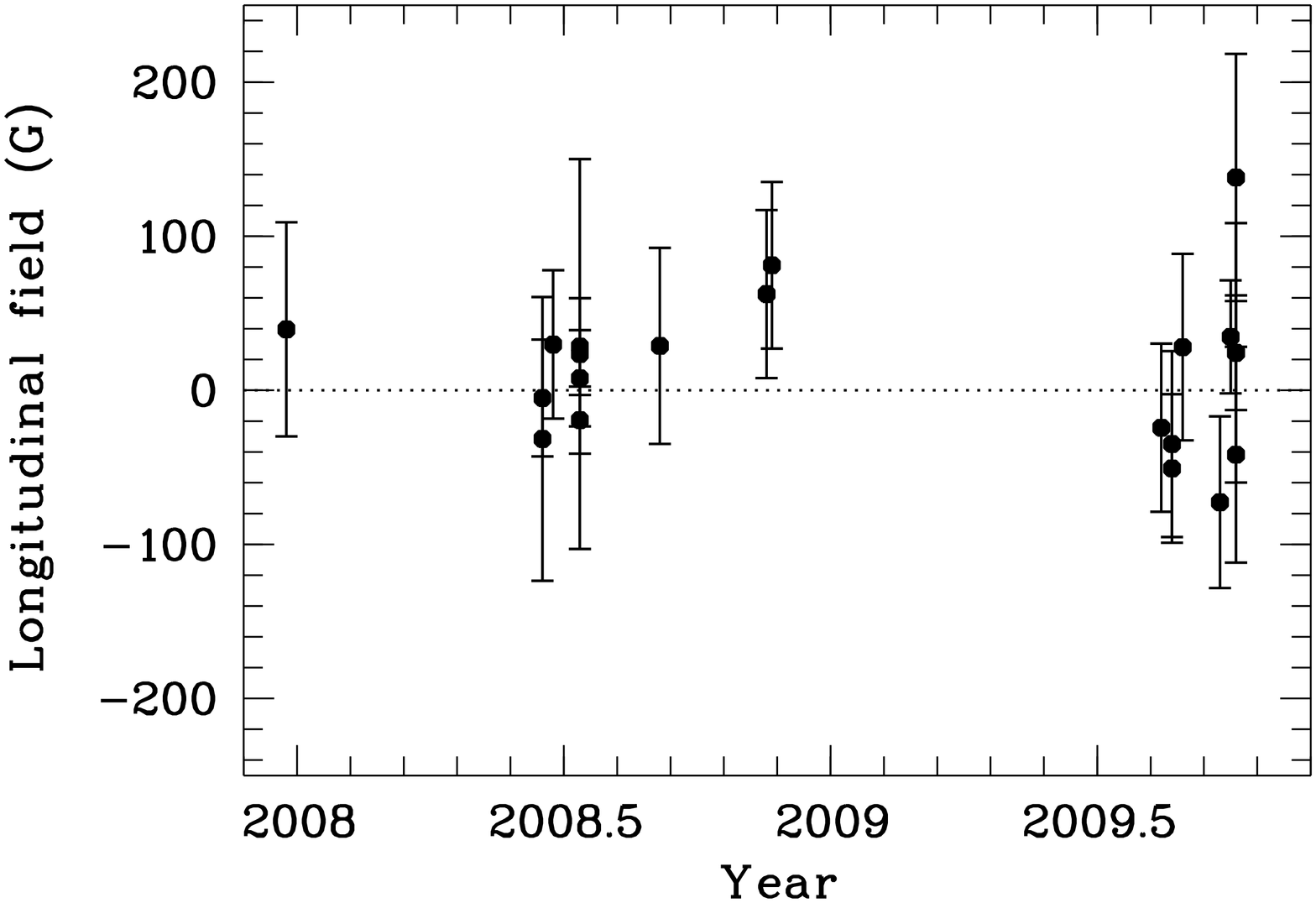}
\caption{{\em dimaPol} magnetic field measurements obtained from H$\beta$
observations of the magnetic null standard $\iota$\,Peg over an approximately two year period.
The scatter of the individual measurements around 0\,G is in good agreement with the mean size
of the individual error bars, confirming that the error bars are reasonable estimates of the 
uncertainties in the observations.
Note that three additional data points have been omitted from the plot; all three were obtained 
on a cloudy night and have large error bars (within $3\sigma$ of 0\,G) that make inclusion in the 
plot difficult.
\label{hd210027}}
\end{figure}

One null standard, $\iota$\,Peg, has been observed 21 times on 15 different nights.
The scatter of the resulting 21 individual H$\beta$ longitudinal magnetic field measurements about
the 0\,G level is 48\,G, in good agreement with the average $\sigma_{B_l} = 58$\,G for the individual
observations (Figure~\ref{hd210027}).
This again reassures us that our derived error bars adequately reflect the true accuracy of the
observations.

As an additional test of {\em dimaPol} performance we
have  carried out observations of scattered solar light during several observing runs.
For such observations the telescope is usually pointed at the sky near the zenith during the day
and long series of short exposures (typically 4 to 6\,s, although as long as 30\,s on several occasions)
are obtained.
Very high precision measurements can be acquired because of the brightness
of the light source and the high contrast provided by the solar spectrum's narrow lines.
We have carried out these observations 37 times over a period
of two years and a total of 3660 individual spectra have been obtained.
No systematic shifts between the individual RO+RE and LO+LE spectra have been detected at a
level corresponding to a longitudinal magnetic field strength of 15\,G.

\begin{figure}[b!]
\includegraphics[angle=270,scale=.50]{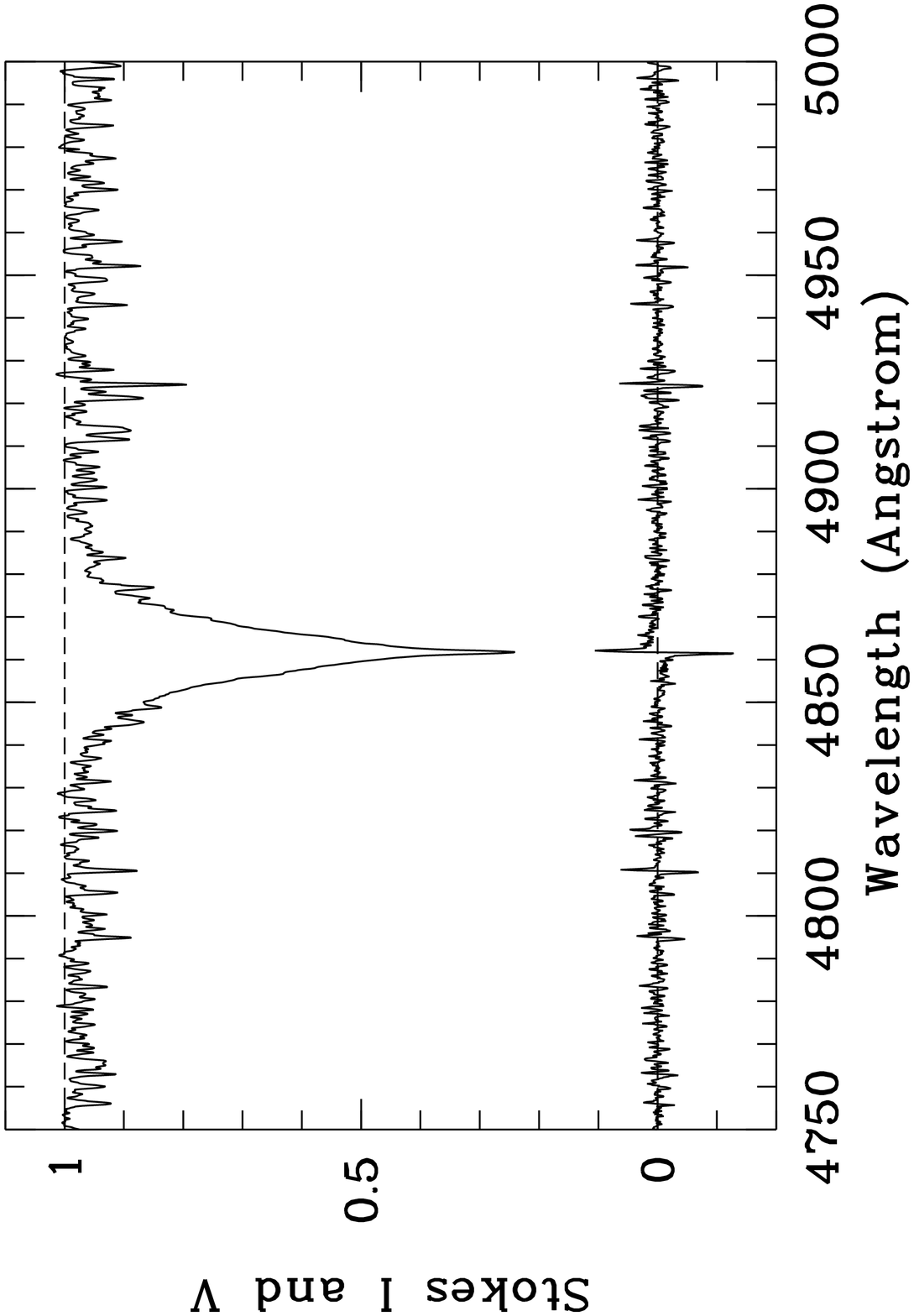}
\caption{A typical {\em dimaPol} observation of the prototypical magnetic
Ap star $\alpha^2$\,CVn.
The Stokes $I$ spectrum is shown at the top of the figure while the lower spectrum is the Stokes
$V$ signature multiplied by a factor of 10 relative to the $I$ spectrum.   
The total exposure time of the observation was 15\,min and the longitudinal
magnetic field strength derived from the H$\beta$ line is -965 $\pm$ 41\,G.
\label{hd112413-stokesIV}}
\end{figure}

Our discussion above shows that {\em dimaPol} does not produce spurious 
magnetic field detections at least at the level of a few tens of G. 
This does not, however, guarantee that the instrument can reliably measure a real magnetic signal 
when a field is present.
In order to regularly test the instrument's ability to detect and accurately evaluate longitudinal magnetic
fields we also attempt to make observations of a few well-known magnetic Ap stars during each
observing run.
One example of an object which we have repeatedly observed is the bright, prototypical magnetic Ap star
HD\,112413 ($\alpha^2$\,CVn).
The star has a well-established rotation period of $5.^{\!\!\rm{d}}46939$ \citep{farnsworth32} and
a strong, reversing longitudinal magnetic field that varies between -1 and +1\,kG \citep{kochukhov10}.
At the time of writing, we have observed it 34 times with
total integration times ranging from 7\,min to one hour, with an average of 20\,min.
A representative observation showing the magnetic signature in the Stokes $V$ profile is
illustrated in Figure~\ref{hd112413-stokesIV}.
Note that the scale used for the Stokes $V$ spectrum in this figure is the same as that used in
Figure~\ref{vega-stokesIV} for Vega; the signature of the magnetic field in $\alpha^2$\,CVn is obvious.

Longitudinal magnetic field measurements of $\alpha^2$\,CVn have been obtained by using 
both the H$\beta$ line and the single metallic line \ion{Fe}{2} $\lambda$4923.
The latter feature was selected because it is the strongest metallic line in the wavelength range
covered by {\em dimaPol} and, with a Land\'{e} value of 1.69, it is also quite sensitive
to the presence of a magnetic field.
Table~\ref{magnetic_data} provides the details of the magnetic field observations including 
the HJD at the mid-point of the spectropolarimeter data acquisition, 
the longitudinal magnetic field values derived from both spectral lines, and the rotation phase derived from the ephemeris of \cite{farnsworth32}: $\rm{JD}=2419869.720 +  5.46939 \times \rm{E}. \label{ephem}$

\begin{figure}[t!]
\includegraphics[angle=270,scale=.50]{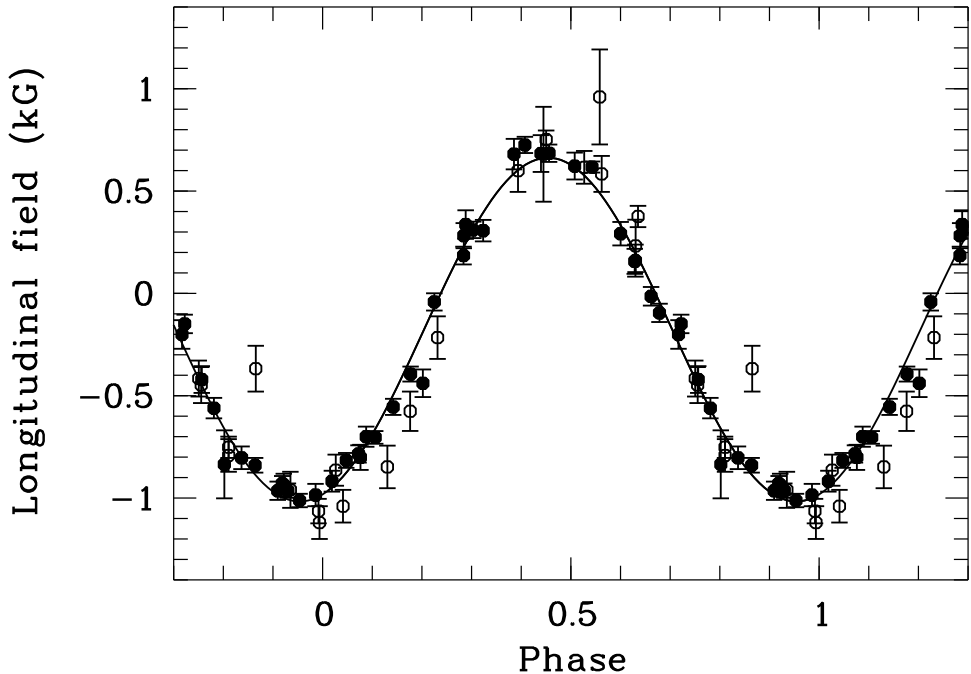}
\caption{A comparison of longitudinal magnetic field observations of
$\alpha^2$\,CVn obtained from H$\beta$ line observations and phased on its $5.^{\!\!\rm{d}}46939$ rotation period. The solid points
are new observations obtained with {\em dimaPol} while
the open points are from \cite{borra77} after correction for the Stark effect (see text). The solid curve is a best-fit sinusoid to the DAO data.
\label{hd112413_hydrogen}}
\end{figure}

\begin{figure}[t]
\includegraphics[angle=270,scale=.50]{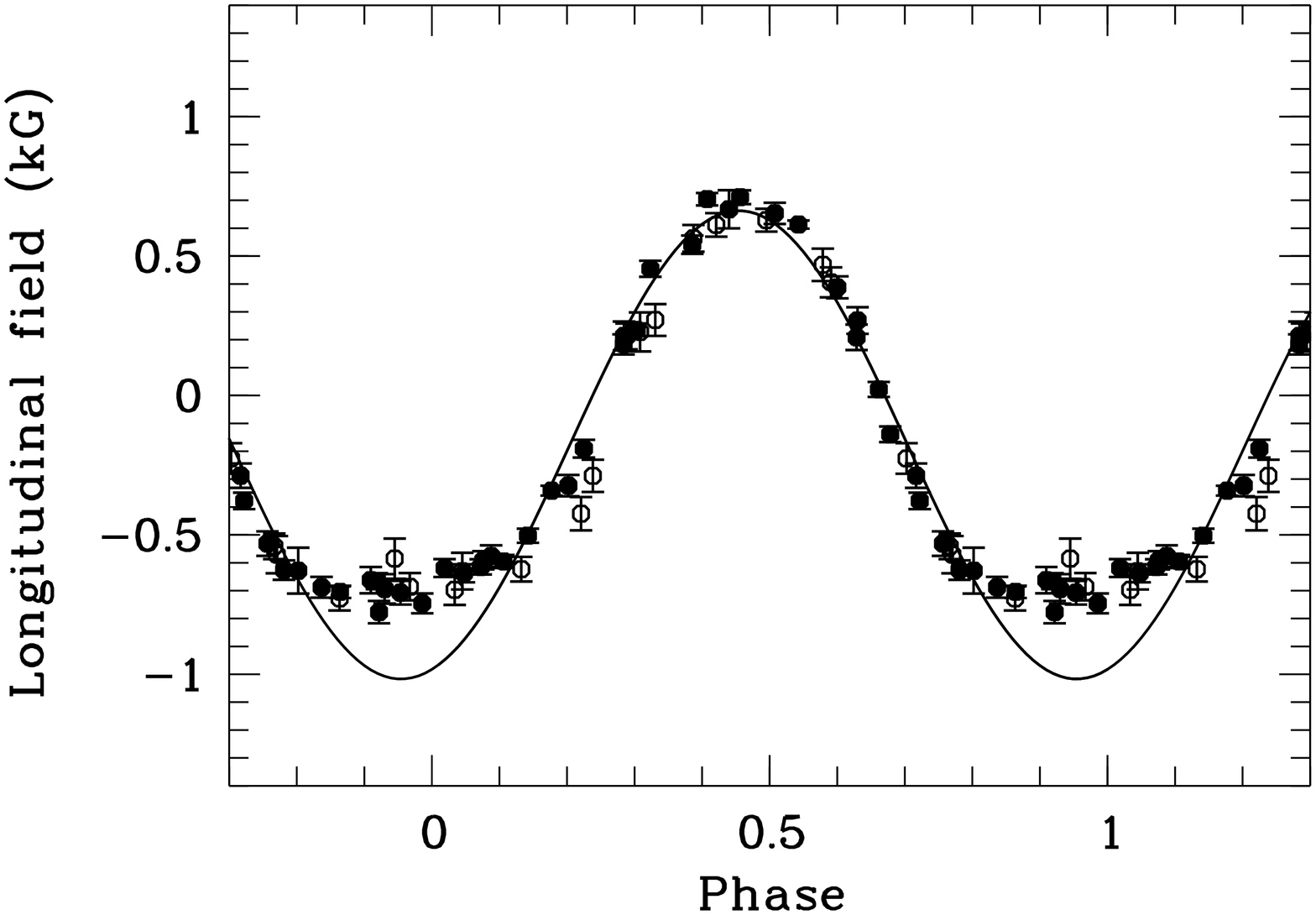}
\caption{As in Figure~\ref{hd112413_hydrogen} but for metallic line observations.
The solid points are new observations obtained with
{\em dimaPol} from the single \ion{Fe}{2} $\lambda$4923 line while the open points 
are measurements derived from LSD of multiple metallic lines \citep{wade00} observed with
MuSiCoS.
The best sinusoidal fit to the DAO H$\beta$ measurements (the same as in
Figure~\ref{hd112413_hydrogen}) is also shown.
Note a significant difference between the H$\beta$ fit and the metallic line
measurements near the negative extremum (see text).
\label{hd112413_metal}}
\end{figure}

The resulting magnetic field curves produced from the measurements of the H$\beta$ and
\ion{Fe}{2} lines are shown in Figures~\ref{hd112413_hydrogen} and \ref{hd112413_metal}
respectively.
Figure~\ref{hd112413_hydrogen} compares {\em dimaPol} H$\beta$
longitudinal magnetic field measurements to those published by \cite{borra77} and
obtained with a two-channel Balmer-line photoelectric Pockels cell polarimeter.
The Pockels cell polarimeter measurements have been multiplied by $4/5$
as suggested by \citet{mathys00} in order to account for the Stark
effect. This correction only applies to measurements in the line wings where
the Stark effect is a factor.
The agreement between the two sets of observations is very good. 
Note that the error bars for our new DAO data are, on average, almost two times better (56\,G versus 98\,G) than 
those of  \cite{borra77}  despite the approximately
three times shorter integration time for the new observations.
The scatter of the DAO points around a best fit sinusoid is 60\,G, quite
close to the average error bar of our measurements.
This is another indication that our error bars adequately represent
the accuracy of our magnetic field measurements.

Figure~\ref{hd112413_metal} shows our longitudinal magnetic field measurements obtained
with the \ion{Fe}{2} $\lambda$4923 line as well as values derived from the Least Squares
Deconvolution (LSD) of multiple metallic lines \citep{wade00} using spectra obtained
with the MuSiCoS spectropolarimeter at Pic du Midi.
Again, the two sets of measurements are in very good agreement.
We also again point out the fact that the average size of the error bars of our single-line
magnetic field measurements (40\,G) is slightly better than that
of the multi-line LSD measurements (55\,G) obtained with a comparable telescope. 

As noted earlier, one of the science goals for which we designed {\em dimaPol}
was an investigation of the occasional differences observed in the longitudinal magnetic field
measured for a single star with the H$\beta$ line versus metallic lines.  
Comparison of Figures~\ref{hd112413_hydrogen} and \ref{hd112413_metal} clearly illustrates such a
difference for magnetic field measurements of $\alpha^2$\,CVn and a detailed discussion of
this (occasionally very large) effect in this and other stars, 
and its possible origin, will be presented in a future paper.

\section{Conclusion}

We have designed and constructed a very inexpensive dual-beam polarimeter for use on the DAO 1.8-m Plaskett Telescope.
The polarimeter, {\em dimaPol}, is implemented as a plug-in module for the telescope's Cassegrain
spectrograph and provides medium resolution circular spectropolarimetry of point sources
over a 260\,\AA\ wide spectral window centered on the H$\beta$ line.
The instrument therefore provides DAO 1.8-m observers the ability to measure longitudinal magnetic fields using the Zeeman effect in  both the H$\beta$ and metal lines of the peculiar A and B stars of the upper main sequence.

We demonstrate that the use of an electrically switchable ferro-electric liquid crystal (FLC) half-wave plate to quickly switch the orthogonal polarizations between the instrument's two channels, in combination with synchronized shuffling of the accumulated charge on the CCD, greatly reduces instrumental effects and increases the accuracy of the  magnetic field measurements.
With extensive on-sky tests we show that by executing 30 to 100 such switches
the accuracy of the longitudinal magnetic field measurements can be improved
by a factor of two.
The use of the FLC wave plate also eliminates the need for any moving parts in the polarimeter module and therefore provides better stability, reliability and significantly lower construction and maintenance costs.

A near-real-time data reduction pipeline has been implemented with MIDAS and provides observers with the ability to obtain a longitudinal magnetic field measurement and its uncertainty within seconds of completing an observing sequence on a star.
The observed shift between the recorded left- and right-circularly polarized spectra is performed with a Fourier cross-correlation technique that is not sensitive to continuum variations or intrinsic spectral line shapes and so can be carried out without interaction from the user.
For magnetic field measurements using the H$\beta$ line {\em dimaPol} provides the simple relation $B_l = 6.8 \times \Delta X$\,kG where $\Delta X$ is the observed shift between the polarized spectra in pixels.

A number of bright, well-established magnetic stars as well as magnetic `null' standards have been repeatedly observed over a two-year period. 
No systematic errors have been found at the level of 15 to 20\,G.
Our multiple observations of the magnetic Ap standard $\alpha^2$\,CVn (as well as other objects not discussed here) clearly
demonstrate that the  longitudinal magnetic field derived from {\em dimaPol}
H$\beta$ and metal line observations agree well with previous published results.
In addition, the derived uncertainties of the field measurements compare favorably with or exceed the accuracies achieved with other often much more expensive polarimeters on similar sized telescopes.
To provide a rough estimate of the capabilities of {\em dimaPol}, observations of a typical sixth or seventh magnitude star in our current
science programs reach an uncertainty of 100 to 200\,G in a typical one to two hour observation
\citep[e.g., see][]{bohlender11}.

We have begun several research programs with {\em dimaPol} on the DAO 1.8-m telescope but encourage any interested users to apply for observing time with the instrument.
Our main survey program consists of an extensive search for magnetic fields in previously poorly studied (and generally relatively faint, $V>7$) Ap and Bp stars.
In a second program we are taking advantage of the capability of the instrument to provide longitudinal
magnetic field measurements in both the H$\beta$ line as well as metal lines to investigate the reality and cause of the often very discrepant magnetic field strengths measured with the two diagnostics, but frequently with very different instrumentation.
As discussed in the previous section, $\alpha^2$\,CVn is an example of one such object.

Despite its cost of only several thousand dollars, {\em dimaPol} is proving to be a very capable addition to the complement of instrumentation available on the DAO 1.8-m Plaskett Telescope and has opened up an entirely new avenue of research at the DAO.
At such a low cost many institutes with spectroscopic capabilities on even very modest telescopes could readily consider building a similar device for research purposes or student training.  

\acknowledgments

We would like to thank J.D.\ Landstreet, G.A.\ Wade, G.A.\ Chuntonov, and
V.G.\ Shtol' for very valuable discussions. J.H.\ Grunhut also provided
valuable help during the first {\em dimaPol} commissioning runs.

{\it Facilities:} \facility{DAO:1.85m}.

\begin{deluxetable}{rrrr}
\tabletypesize{\scriptsize}
%\rotate
\tablecaption{Longitudinal magnetic field measurements for the Ap star
$\alpha^2$\,CVn obtained with {\em dimaPol}.
\label{magnetic_data}}
\tablewidth{0pt}
\tablehead{
\colhead{HJD} & $B_l$ (G) & $B_l$ (G) & \colhead{$\phi$} \\
\colhead{(-245\,0000)} & (H$\beta$) & (Fe\,II) & 
}
\startdata
4480.98881   & $ -394 \pm \ 41$ & $ -342 \pm \ 19$ & 0.1771 \\
4487.04114   & $  185 \pm \ 54$ & $  183 \pm \ 35$ & 0.2837 \\
4487.98377   & $  685 \pm \ 45$ & $  711 \pm \ 26$ & 0.4561 \\
4543.89436   & $  -95 \pm \ 54$ & $ -140 \pm \ 34$ & 0.6785 \\
4632.77958   & $ -965 \pm \ 41$ & $ -694 \pm \ 62$ & 0.9299 \\
4633.74205   & $ -704 \pm \ 34$ & $ -596 \pm \ 27$ & 0.1059 \\
4634.71760   & $  282 \pm \ 61$ & $  213 \pm \ 57$ & 0.2843 \\
4634.73870   & $  337 \pm \ 69$ & $  211 \pm \ 47$ & 0.2881 \\
4636.78101   & $  -14 \pm \ 50$ & $   21 \pm \ 27$ & 0.6615 \\
4640.74014   & $  681 \pm \ 76$ & $  539 \pm \ 34$ & 0.3854 \\
4659.76615   & $ -841 \pm \ 39$ & $ -706 \pm \ 23$ & 0.8640 \\
4660.77011   & $ -817 \pm \ 41$ & $ -641 \pm \ 35$ & 0.0476 \\
4661.74017   & $  -41 \pm \ 50$ & $ -195 \pm \ 39$ & 0.2250 \\
4870.00537   & $  311 \pm \ 40$ & $  236 \pm \ 22$ & 0.3033 \\
4877.95372   & $ -421 \pm \ 64$ & $ -531 \pm \ 47$ & 0.7565 \\
4878.08744   & $ -561 \pm \ 48$ & $ -627 \pm \ 34$ & 0.7810 \\
4878.85740   & $ -937 \pm \ 60$ & $ -778 \pm \ 42$ & 0.9217 \\
4879.03109   & $-1013 \pm \ 33$ & $ -708 \pm \ 42$ & 0.9535 \\
4902.93131   & $  307 \pm \ 50$ & $  454 \pm \ 30$ & 0.3233 \\
4903.93935   & $  623 \pm \ 72$ & $  651 \pm \ 40$ & 0.5076 \\
4934.75652   & $ -555 \pm \ 47$ & $ -507 \pm \ 34$ & 0.1421 \\
4936.94704   & $  618 \pm \ 28$ & $  614 \pm \ 16$ & 0.5426 \\
4937.92873   & $ -149 \pm \ 50$ & $ -379 \pm \ 32$ & 0.7221 \\
4939.84380   & $ -785 \pm \ 44$ & $ -616 \pm \ 27$ & 0.0722 \\
4939.86512   & $ -802 \pm \ 59$ & $ -591 \pm \ 39$ & 0.0761 \\
4959.77708   & $ -201 \pm \ 68$ & $ -287 \pm \ 45$ & 0.7168 \\
4960.88101   & $ -930 \pm \ 37$ & $ -666 \pm \ 30$ & 0.9186 \\
4963.73155   & $  684 \pm \ 99$ & $  668 \pm \ 67$ & 0.4398 \\
4964.76458   & $  157 \pm \ 63$ & $  209 \pm \ 52$ & 0.6287 \\
5058.69226   & $ -836 \pm  141$ & $ -628 \pm \ 84$ & 0.8020 \\
5059.69697   & $ -986 \pm \ 65$ & $ -746 \pm \ 50$ & 0.9857 \\
5068.69036   & $  160 \pm  100$ & $  269 \pm \ 61$ & 0.6300 \\
5321.81108   & $ -965 \pm \ 47$ & $ -662 \pm \ 50$ & 0.9095 \\
5322.78558   & $ -701 \pm \ 51$ & $ -576 \pm \ 39$ & 0.0877 \\
\enddata
\end{deluxetable}


\begin{thebibliography}{}

\bibitem[Bagnulo et al.(2002)]{bagnulo02} Bagnulo, S., Szeifert, T.,
 Wade, G.A., Landstreet, J.D., \& Mathys, G. 2002, \aap, 389, 191

\bibitem[Bland-Hawthorn \& Barton(1994)]{bland95} Bland-Hawthorn, J., \&
 Barton, J. 1995, AAO Newsl., 74, 10

\bibitem[Blinov \& Chigrinov(1994)]{blinov94} Blinov, L., \& Chigrinov, V.
 1994, Electrooptic Effects in Liquid Crystal Materials
 (New York, Springer-Verlag), 372
    
\bibitem[Bohlender \& Monin(2011)]{bohlender11} Bohlender, D.A., \& Monin, D.
 2011, \aj, 141, 169

\bibitem[Borra \& Landstreet(1973)]{borra73} Borra, E.F., \& Landstreet, J.D.
 1973, \apj, 185, L139

\bibitem[Borra \& Landstreet(1977)]{borra77} Borra, E.F., \& Landstreet, J.D.
 1977, \apj, 212, 141

\bibitem[Borra \& Landstreet(1980)]{borra80} Borra, E.F., \& Landstreet, J.D.
 1980, \apjs, 42, 421

\bibitem[Bray \& Loughhead(1964)]{bray64} Bray, R.J., \& Loughhead, R.D. 1964,
 Sunspots (London, Chapman and Hall), chap. 5

\bibitem[Bychkov, Bychkova, \& Madej(1977)]{bychkov05} Bychkov, V.D.,
Bychkova L.V., \& Madej, J. 2005, \aap, 430, 1143
    
\bibitem[Casini \& Landi Degl'Innocenti(1994)]{casini94} Casini, R., \&
 Landi Degl'Innocenti, E. 1994, \aap, 291, 668

\bibitem[Donati et al.(1997)]{donati97} Donati, J.-F., Semel, M.,
 Carter, B.D., Rees, D.E., \& Collier Cameron, A. 1997, \mnras, 291, 658

\bibitem[Donati \& Landstreet(2009)]{donati09} Donati, J.-F., \&
 Landstreet, J.D. 2009, \araa, 47, 333

\bibitem[Farnsworth(1932)]{farnsworth32} Farnsworth, G. 1932, \apj, 76, 313

\bibitem[Gisler et al.(2004)]{gisler04} Gisler, D., Schmid, H.M.,
 Thalmann, C., Povel, H.P., Stenflo, J.O., Joos, F., Feldt, M., Lenzen, R.,
 Tinbergen, J., Gratton, R., Stuik, R., Stam, D.M., Brandner, W., Hippler, S.,
 Turatto, M., Neuhauser, R., Dominik, C., Hatzes, A., Henning, T., Lima, J.,
 Quirrenbach, A., Waters, L.B.F.M., Wuchterl, G., \& Zinnecker, H.
 2004, \procspie, 5492, 463

\bibitem[Harrington et al.(2010)]{harrington10} Harrington, D.M., Kuhn, J.R., Sennhauser, C., Messersmith, E.J., Thornton, R.J. 2010, \pasp, 122, 420

\bibitem[Keller et al.(2003)]{keller03} Keller, C. U., Harvey, J. W., \&
 The Solis Team 2003, in proceedings of ASP Conference Proceedings:
 Solar Polarization, ed. Javier Trujillo-Bueno and Jorge Sanchez Almeida,
 (San Francisco, Astronomical Society of the Pacific), 307, 13

\bibitem[Kochukhov \& Wade(2010)]{kochukhov10} Kochukhov, O., \& Wade, G.A.
 2010, \aap, 513, A13

\bibitem[Kupka et al.(2000)]{kupka00} Kupka, F., Ryabchikova, T.A.,
 Piskunov, N.E., Stempels, H.C., \& Weiss, W.W. 2000, Baltic Astronomy, 9, 590

\bibitem[Landstreet(1992)]{landstreet92} Landstreet, J. D. 1992, \aapr, 4, 35

\bibitem[Lignieres et al.(2009)]{lignieres09} Lignieres, F., Petit, P.,
 Bohm, T., \& Auriere, M. 2009, \aap, 500, 41L

\bibitem[Malherbe et al.(2004)]{malherbe04} Malherbe, J.-M., Roudier, Th.,
 Mein, P., Moity, J., \& Muller, R. 2004, \aap, 427, 745

\bibitem[Mart\'{i}nez Pillet et al.(1999)]{martinez99} Mart\'{i}nez Pillet, V.,
 Collados, M., S\'{a}nchez Almeida, J., Gonz\'{a}lez, V., Cruz-Lopez, A.,
 Manescau, A., Joven, E., Paez, E., Diaz, J., Feeney, O., S\'{a}nchez, V.,
 Scharmer, G., \& Soltau, D. 1999, High Resolution Solar Physics: Theory,
 Observations, and Techniques, ASP Conference Series, 183,
 Eds. Rimmele, T.R., Balasubramaniam, K.S., \& Radick, R.R., 264

\bibitem[Mathys(1988)]{mathys88} Mathys, G. 1988, \aap, 189, 179

\bibitem[Mathys et al.(2000)]{mathys00} Mathys, G., Stehl\'{e}, C.,
 Brillant, S., \& Lanz, T. 2000, \aap, 358, 1151

\bibitem[McLean et al.(1981)]{mclean81} McLean, I.S., Cormack, W.A.,
 Herd, J.T., \& Aspin, C. 1981, \procspie, 290, 155

\bibitem[Monin, Fabrika, \& Valyavin(2002)]{monin02} Monin, D.N.,
 Fabrika, S.N., \& Valyavin, G.G. 2002, \aap, 396, 131
 
 \bibitem[Petit et al.(2010)]{petit10} Petit, P., Lignieres, F., Wade, G.A.,
 Auriere, M., Boehm, T., Bagnulo, S., Dintrans, B., Fumel, A., Grunhut, J.,
 Lanoux, J., Morgenthaler, A., \& van Groootel, V., \aap, 523, 41

\bibitem[Povel (2001)]{povel01} Povel, H.P.
 2001, in proceedings of ASP Conference Proceedings:
 Magnetic Fields Across the Hertzsprung-Russell Diagram,
 ed. G. Mathys, S.K. Solanki, and D.T. Wickramasinghe,
 (San Francisco, Astronomical Society of the Pacific), 248, 543

\bibitem[Rees \& Semel (1979)]{rees79} Rees, D.E., \& Semel, M.D. 1979,
 \aap, 74, 1

\bibitem[Richardson(1968)]{richardson68} Richardson, E.H. 1968, \jrasc, 62, 313

\bibitem[Wade et al.(2000)]{wade00} Wade, G.A., Donati, J.-F.,
 Landstreet, J.D., \& Shorlin, S.L.S. 2000, \mnras, 313, 851

\end{thebibliography}
\end{document}